\documentclass{article}
\usepackage[margin=30mm]{geometry}
\usepackage{graphicx} %
\usepackage{rotating} %
\usepackage[
    style=authoryear, 
    backend=biber, 
    natbib=true,
    date=year,
    isbn=false,
    url=false,
    doi=false,
    eprint=false
]{biblatex}
\renewbibmacro{in:}{}

\usepackage{amsmath, amssymb, amsthm, accents}
\usepackage{bm}
\usepackage{booktabs}
\usepackage{setspace}
\begin{document}
\title{Estimating Rail Commuter Preferences under Endogenous Crowding: A Population Game Approach to Train Choice in Tokyo\thanks{Shota Fujishima acknowledges financial support from JSPS KAKENHI Grant Numbers JP26K00334, JP23K22098, and JP24K03007.}}
\author{Shota Fujishima$^{a}$ \hspace{3em} Ryosuke Matsumoto$^{a,b,\dagger}$\\[0.2em]
{\small $^{a}$Graduate School of Economics, Hitotsubashi University}\\
{\small $^{b}$Information and Communication Technology Division, Railway
Technical Research Institute}}
\date{}
\maketitle
\begingroup
\renewcommand{\thefootnote}{$\dagger$}
\footnotetext{Corresponding author:
\texttt{matsumoto.ryosuke.84@rtri.or.jp}.}
\endgroup

\begin{abstract}
We estimate commuters' preferences over schedule delay, travel time, and crowding from observed train choices on a commuter rail line in Tokyo. Because crowding is itself an outcome of these choices, we model train choice as a game in which train choices and crowding are jointly determined, and estimate the preferences by maximum likelihood using the closed-form stationary distribution of the game. Using individual-level data from Japan's 2015 Public Transport Survey for Metropolitan Areas matched with train timetables, we construct desired arrival times from reported work start times rather than inferring them from observed arrival times. We find that schedule delay costs are large relative to crowding costs. One minute of schedule delay early and late corresponds to 0.39 and 0.49 minutes of in-vehicle travel time, respectively, whereas one minute of travel at a crowding rate of 100\% adds a cost equivalent to 0.04 minutes. The estimated crowding cost is, however, sensitive to the specification of desired arrival times. These results suggest that scheduling constraints are central to evaluating policies to reduce peak-hour crowding.


\end{abstract}

\section{Introduction}
\label{sec:introduction}

Why do so many commuters choose crowded trains during the morning peak hours? Crowding imposes disutility on commuters through physical discomfort from close contact with other passengers, the burden of standing, and limited opportunities to use travel time productively. \citet{internationaltransportforum2014} identifies crowding as an important component of the generalized cost of public transport, and reviews of the literature on crowding valuation find that crowding substantially raises the perceived cost of in-vehicle travel time \citep{wardman2011TransportReviews,li2011TransportPolicy}. To reduce crowding, policies that encourage commuters to travel outside the peak, such as time-dependent fares and off-peak commuting incentives, have been implemented or proposed alongside capacity expansion \citep{depalma2015TransportationResearchPartB:Methodological,internationaltransportforum2021}. The effectiveness of such policies depends on whether commuters are willing to change the trains they take. Work start times constrain this adjustment because a less crowded train may arrive well before work begins or make the commuter late. It may also be a local train that stops at more stations and therefore takes longer to reach the destination. Commuters must therefore weigh the benefits of avoiding crowding against the costs of changes in arrival time and travel time.

This paper estimates commuters' preferences over schedule delay, travel time, and crowding from their observed train choices in urban rail. This requires addressing two related challenges. First, crowding is itself an outcome of the train choices we seek to explain. Second, observed arrival times reflect commuters' responses to crowding and travel time as well as their scheduling preferences, so they need not coincide with desired arrival times. We address these challenges by combining a model in which train choice and crowding are jointly determined with survey information on work start times used to construct desired arrival times. 

We model urban rail commuting by combining the standard trip-scheduling framework \citep{small1982Am.Econ.Rev.} with crowding costs, following the urban rail commuting literature \citep{depalma2017JournalofUrbanEconomics}. Each commuter chooses one train from those available at their origin station to minimize the sum of four cost components: schedule delay early, schedule delay late, travel time, and crowding. The crowding cost of each train depends on the number of commuters choosing it, together with the number of other passengers, such as those traveling for other purposes or to other destinations, and its capacity.

We formulate the interaction between train choice and crowding as a population game \citep{sandholm2010}. A population game describes strategic interaction among a large number of anonymous players. Each player's payoff depends only on their own strategy and on the distribution of strategies across the population, not on who chooses which strategy. This structure fits urban rail commuting. A commuter's crowding cost depends on how many commuters choose each train, not on their identities, and each commuter's effect on crowding is small even though aggregate choices determine crowding levels. Commuters at the same origin station face the same set of trains and form one population, so the state of the game is the distribution of commuters across trains at each station.

Our game also has a property that makes estimation tractable: it is a potential game \citep{monderer1996GamesEcon.Behav.}. In a potential game, there is a single function of the distribution of commuters, called the potential function, that increases whenever a commuter switches to a train with a lower cost. We assume that commuters revise their train choices according to a logit choice rule. Under this revision process, the potential-game structure yields a closed-form stationary distribution of commuter allocations. We estimate preference parameters by maximizing a large-population approximation to the log likelihood derived from this distribution. We obtain standard errors and confidence intervals using a parametric bootstrap based on the estimated revision process.

We combine this framework with individual-level cross-sectional data from Japan's 2015 Public Transport Survey for Metropolitan Areas (PTSMA), a survey of public transport users in Japan's major metropolitan areas conducted by the Ministry of Land, Infrastructure, Transport and Tourism (MLIT). The survey data contain information on passengers' boarding and alighting stations and times, trip purposes, and work start times. By matching boarding and alighting information with train timetables, we can infer the trains used by commuters. We construct desired arrival times from reported work start times and onward travel times, together with an assumed preferred buffer time. These data allow us to estimate preferences while accounting for endogenous crowding without inferring desired arrival times from the observed distribution of train choices.

We apply this approach to the Seibu Shinjuku Line, a commuter rail line connecting suburban areas with central Tokyo (Figure~\ref{fig:study-area}). This setting has three features that suit our analysis. First, trains on the line are severely crowded during the morning peak. Second, most commuters on the line alight at one of the two adjacent stations at its central-Tokyo end, so relatively few commuters get off at intermediate stations, for example to transfer to other lines. This limits the part of crowding that our model must treat as exogenous. Third, the line operates express and local trains that differ in arrival time, travel time, and crowding. These differences, together with information on commuters' desired arrival times, allow us to distinguish the four cost components. Section~\ref{sec:background} describes the line in detail.

We find that, for urban rail commuters, schedule delay costs are large relative to crowding costs. In the baseline estimation, all four cost parameters for schedule delay early, schedule delay late, travel time, and crowding are positive, and their 95\% confidence intervals from the parametric bootstrap exclude zero. We express these costs in travel-time equivalents using the estimated in-vehicle travel-time coefficient, excluding the additional cost of crowding. One minute of schedule delay early corresponds to 0.39 minutes of in-vehicle travel time, while one minute of schedule delay late corresponds to 0.49 minutes. One minute of travel at a crowding rate of 100\% imposes an additional crowding cost equivalent to 0.04 minutes of in-vehicle travel time. These findings suggest that commuters may choose crowded trains to avoid arriving earlier or later than desired, despite the discomfort caused by crowding. The relatively large schedule delay costs compared with crowding costs also remain evident when we allow for heterogeneity in desired arrival times. However, the estimated crowding parameter is sensitive to the specification of desired arrival times and reaches zero in some specifications.

The estimated preferences highlight the importance of commuters' scheduling constraints when evaluating policies to reduce peak-hour crowding. Shifting to a less crowded train may entail costs from arriving earlier or later than desired, which need to be weighed against the benefits of reduced crowding and any incentives offered, such as off-peak fare discounts and reward points. Whereas such incentives leave desired arrival times unchanged, measures that relax scheduling constraints, such as staggered work start times and flexible working hours, may make such adjustments less costly. This suggests a role for coordination between employers and transport providers when designing policies to spread commuting demand across time. Whether these measures complement fare-based incentives, and how much they reduce crowding, remain questions for counterfactual analysis.\footnote{Counterfactual simulations of specific policies are part of our ongoing work and will be included in a future version of this paper.}

This study makes three contributions, which relate to the economics of trip scheduling, the valuation of crowding, and the estimation of transport equilibrium models, respectively. First, we extend the empirical analysis of trip scheduling to urban rail. Since \citet{small1982Am.Econ.Rev.}, the trip-scheduling literature has studied how commuters trade off travel time against deviations from their desired arrival times. This literature has focused mainly on road transport, where traveling at the peak is costly because congestion lengthens travel times. \citet{kreindler2024ECTA} is a recent example that estimates scheduling preferences within an equilibrium model with endogenous road congestion. In urban rail, by contrast, trains run on a timetable, so traveling at the peak is costly mainly because of crowding rather than longer travel times. We extend the empirical analysis of trip scheduling to this trade-off between schedule delay and crowding, which has received less attention in public transport \citep{depalma2017JournalofUrbanEconomics}.

Second, this study contributes to the literature on crowding valuation in public transport. Much of this literature relies on stated preference (SP) data, in which respondents choose among hypothetical alternatives that differ in crowding, travel time, and cost \citep{wardman2011TransportReviews,li2011TransportPolicy}. Because these choices are hypothetical, they may not fully reflect actual travel behavior. Studies using revealed preference (RP) data instead infer crowding costs from observed choices, such as those of metro passengers in Singapore who ride in the opposite direction to secure a seat \citep{tirachini2016TransportPolicy} and route choices on the London Underground and on tram and bus networks in the Netherlands \citep{horcher2017TransportationResearchPartB:Methodological,yap2020TransportmetricaA:TransportScience}. These RP studies, however, treat crowding as a given attribute of each alternative rather than as an outcome of passengers' choices. We account for this endogeneity when estimating crowding preferences from observed train choices.

Finally, this study contributes to the structural estimation of urban rail equilibrium models, in terms of both the estimation method and the data. \citet{depalma2017JournalofUrbanEconomics} theoretically analyze an equilibrium model of urban rail commuting with schedule delay and crowding, but do not estimate preference parameters from observed train choices. \citet{xu2024TransportationResearchPartB:Methodological} use smart card data to estimate commuter preferences through inverse optimization, minimizing squared differences between observed boarding flows and those predicted by a stochastic user equilibrium model. Our approach differs in both estimation method and the construction of desired arrival times. Methodologically, we use a likelihood for the observed allocation of commuters across trains.\footnote{A related approach based on the stationary distribution of a stochastic evolutionary model is used by \citet{fujishima2025RIETIDP} to quantify congestion externalities in road networks.} We also report standard errors and confidence intervals from a parametric bootstrap, which are not reported for their inverse-optimization estimates. On the data side, \citet{xu2024TransportationResearchPartB:Methodological} use the peaks of observed arrival-time distributions as proxies for desired arrival times. These observed times may reflect adjustments to crowding and travel time as well as scheduling preferences. We construct desired arrival times using separately recorded work start times and information on onward travel, together with an assumed preferred buffer time. This allows us to anchor desired arrival times in information beyond the train choices that the model seeks to explain.\footnote{The survey also records trip purposes, which allows us to distinguish the commuters included in the model from other passengers, whose contribution to crowding we treat as exogenous. Smart card data typically do not record this information.}

The remainder of this paper is organized as follows. Section~\ref{sec:background} describes the institutional background. Section~\ref{sec:model} presents the population game model. Section~\ref{sec:data} describes the data and their construction. Section~\ref{sec:estimation} explains the estimation method and statistical inference for the preference parameters. Section~\ref{sec:results} presents the estimation results. Section~\ref{sec:discussion} compares the estimates with previous studies and discusses their policy implications. Section~\ref{sec:conclusion} concludes and discusses the limitations of this study.

\section{Background}
\label{sec:background}

This section describes the features of commuting on the Seibu Shinjuku Line that motivate the assumptions of our model. Rail plays a central role in commuting in the Tokyo metropolitan area. During the morning peak, a large number of commuters travel toward central Tokyo within a short period. As a result, passenger numbers vary greatly across trains, creating differences in crowding levels. At the same time, rail services are frequent, so commuters can choose among several trains that serve their origin station and arrive in central Tokyo at different times. In addition, fares depend on the origin and destination stations but not on the train, except for services that require a surcharge.\footnote{On the Seibu Shinjuku Line, such services are limited express trains, which stop at only a few major stations. We exclude them from our analysis. On weekdays in November 2015, seven limited express trains arrived in central Tokyo before 10:30 a.m., but none arrived between about 7:20 and 8:50, when trains run most frequently.}
 Unlike rail systems that charge higher fares during peak hours, fares on the Seibu Shinjuku Line during our study period did not vary by time of day.\footnote{Time-dependent fares have since been introduced on some lines in the Tokyo metropolitan area. For example, East Japan Railway Company (JR East) introduced off-peak commuter passes in 2023. See East Japan Railway Company, \textit{JR East Group Report 2023: Integrated Report}, p.~42, \url{https://www.jreast.co.jp/e/environment/pdf_2023/all.pdf}.} Because trains differ little in monetary cost, train choice is mainly a trade-off among arrival time, travel time, and crowding.

\begin{figure}[!htbp]
\centering
\includegraphics[width=0.90\textwidth]{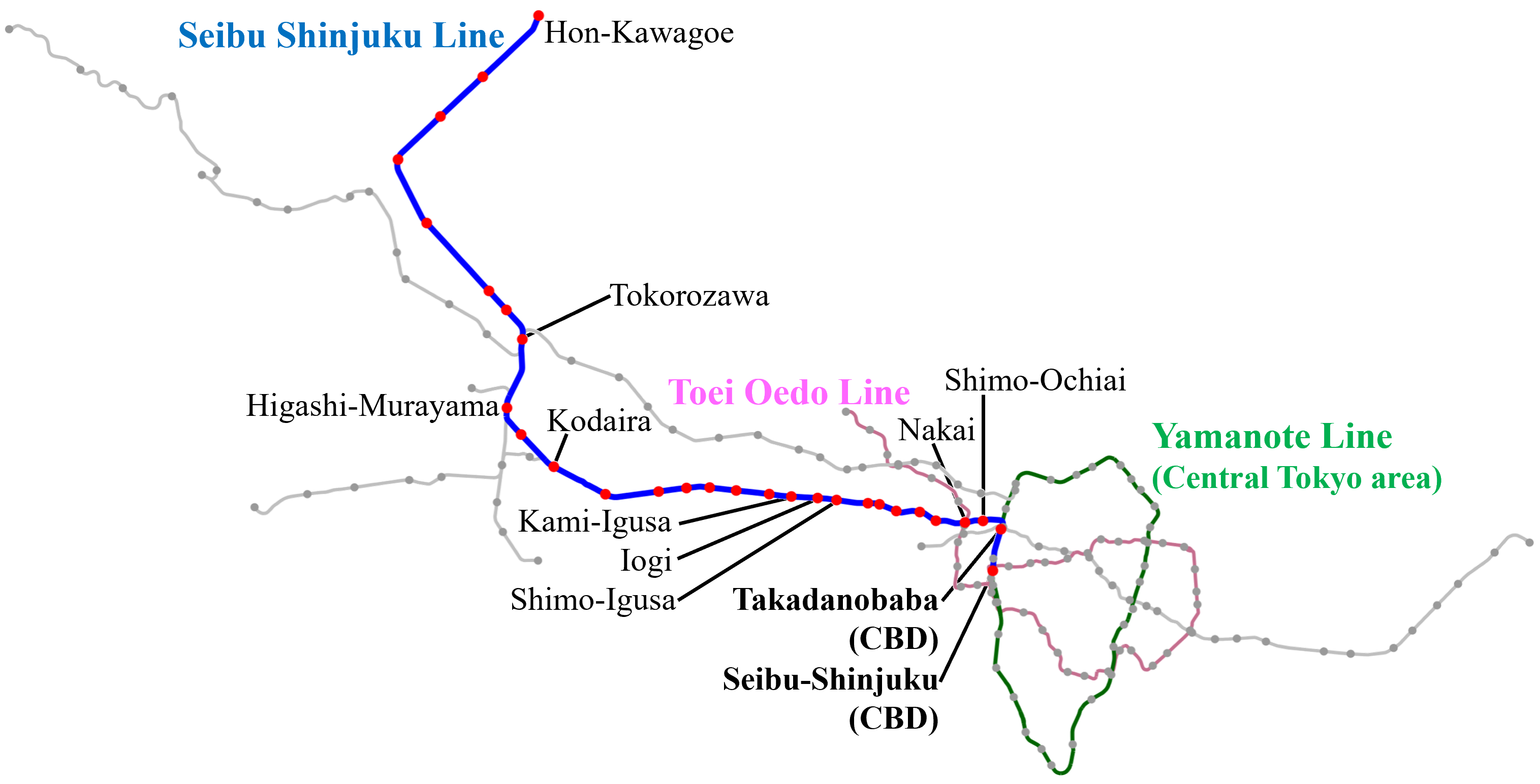}
\caption{The Seibu Shinjuku Line and its connections to other railway lines}
\label{fig:study-area}

\vspace{0.5em}
\begin{minipage}{0.90\textwidth}
\footnotesize
\textit{Notes:} The blue line represents the Seibu Shinjuku Line, while the gray lines represent other railway lines. The green line represents the Yamanote Line, and the pink line represents the Toei Oedo Line. The area within and along the Yamanote Line serves as a rough geographical reference for central Tokyo. Red dots indicate stations on the Seibu Shinjuku Line, while gray dots indicate stations on other lines. The study covers 27 origin stations from Hon-Kawagoe to Shimo-Ochiai and two destination stations, Takadanobaba and Seibu-Shinjuku, which we treat as a single destination (CBD). The map is based on the 2015 National Land Numerical Information railway data provided by MLIT \citep{mlit2015railway}.
\end{minipage}
\end{figure}

Figure~\ref{fig:study-area} shows the route map of the Seibu Shinjuku Line. The line runs from Hon-Kawagoe in the suburbs to Seibu-Shinjuku in central Tokyo. It connects with other railway lines at Tokorozawa, Higashi-Murayama, and Kodaira in the suburbs. Between Kodaira and Takadanobaba, Nakai is the only station that connects with another line, the Toei Oedo Line, and transferring there offers little advantage for reaching central Tokyo over staying on the Seibu Shinjuku Line. As a result, most commuters on the line travel to Takadanobaba or Seibu-Shinjuku, the two adjacent stations at its central-Tokyo end. In the PTSMA data, 80.7\% of commuters riding inbound trains on the line alight at one of these two stations. This structure allows us to analyze commuters' choices among trains on a single line and to treat the remaining passengers as an exogenous contribution to crowding. On a line with many transfer stations, a larger share of the passengers on each train would be exogenous to the model, and explaining their choices would require modeling routes across the rail network as a whole.

In the baseline analysis, each of the 27 stations from Hon-Kawagoe to Shimo-Ochiai is treated as a separate origin zone. We combine Takadanobaba and Seibu-Shinjuku into a single destination, which we refer to as the central business district (CBD). The commuters in our model are those whose trip on the Seibu Shinjuku Line ends at the CBD, including those who transfer there to other lines. Each commuter in the model faces the same decision of which train serving their origin station to take to the CBD. These trains differ in their arrival times at the CBD, travel times, and crowding levels. Table~\ref{tab:train-service-type} shows the stations served by each of the four train types, Commuter Express, Express, Semi-Express, and Local, in order of increasing number of stops, together with the number of trains of each type available at each station. Because stops differ by train type, travel times to the CBD differ across trains, and the set of available trains differs across origin stations.

\begin{table}[!htbp]
\centering
\caption{Train service types and station stops}
\label{tab:train-service-type}
\scriptsize
\setlength{\tabcolsep}{2.5pt}
\begin{tabular}{cccccccccc}
\toprule
 & & \multicolumn{4}{c}{Panel A: Station stops by service type}& \multicolumn{4}{c}{Panel B: Number of trains by service type}\\
ID & Station & Commuter Express& Express& Semi Express& Local & Commuter Express& Express& Semi Express&Local \\
\midrule
1  & Hon-Kawagoe      & \ensuremath{\bullet}&  \ensuremath{\bullet}& \ensuremath{\bullet}& \ensuremath{\bullet} & 2& 15& 7&3\\
2  & Minami-Otsuka    & &  \ensuremath{\bullet}&  \ensuremath{\bullet}& \ensuremath{\bullet} & & 15& 7&3\\
3  & Shin-Sayama      & & \ensuremath{\bullet}&  \ensuremath{\bullet}& \ensuremath{\bullet} & & 15& 7&3\\
4  & Sayamashi        & \ensuremath{\bullet}& \ensuremath{\bullet}&  \ensuremath{\bullet}& \ensuremath{\bullet} & 2& 15& 7&5\\
5  & Iriso            & & \ensuremath{\bullet}&  \ensuremath{\bullet}& \ensuremath{\bullet} & & 15& 7&5\\
6  & Shin-Tokorozawa  & \ensuremath{\bullet}& \ensuremath{\bullet}&  \ensuremath{\bullet}&  \ensuremath{\bullet} & 2& 17& 10&14\\
7  & Kokukoen         & & \ensuremath{\bullet}&  \ensuremath{\bullet}& \ensuremath{\bullet} & & 17& 10&14\\
8  & Tokorozawa       & \ensuremath{\bullet}& \ensuremath{\bullet}&  \ensuremath{\bullet}& \ensuremath{\bullet} & 2& 17& 10&14\\
9  & Higashi-Murayama & \ensuremath{\bullet}& \ensuremath{\bullet}&  \ensuremath{\bullet}& \ensuremath{\bullet} & 2& 17& 10&14\\
10 & Kumegawa         & & \ensuremath{\bullet}&  \ensuremath{\bullet}&  \ensuremath{\bullet} & & 17& 10&14\\
11 & Kodaira          & & \ensuremath{\bullet}&  \ensuremath{\bullet}& \ensuremath{\bullet} & & 27& 16&22\\
12 & Hana-Koganei     & & \ensuremath{\bullet}&  \ensuremath{\bullet}&  \ensuremath{\bullet} & & 27& 16&22\\
13 & Tanashi          & \ensuremath{\bullet}& \ensuremath{\bullet}&  \ensuremath{\bullet}&  \ensuremath{\bullet} & 2& 27& 18&30\\
14 & Seibu-Yagisawa   & & &  \ensuremath{\bullet}& \ensuremath{\bullet} & & & 18&30\\
15 & Higashi-Fushimi  & & &  \ensuremath{\bullet}& \ensuremath{\bullet} & & & 18&30\\
16 & Musashiseki      & & &  \ensuremath{\bullet}& \ensuremath{\bullet} & & & 18&30\\
17 & Kami-Shakujii    & \ensuremath{\bullet}& \ensuremath{\bullet}&  \ensuremath{\bullet}& \ensuremath{\bullet} & 2& 27& 18&42\\
18 & Kami-Igusa       & & &  & \ensuremath{\bullet} & & & &42\\
19 & Iogi             & & &  & \ensuremath{\bullet} & & & &42\\
20 & Shimo-Igusa      & & &  & \ensuremath{\bullet} & & & &42\\
21 & Saginomiya       & \ensuremath{\bullet}& \ensuremath{\bullet}&  \ensuremath{\bullet}& \ensuremath{\bullet} & 2& 27& 18&42\\
22 & Toritsu-Kasei    & & &  & \ensuremath{\bullet} & & & &42\\
23 & Nogata           & & &  & \ensuremath{\bullet} & & & &42\\
24 & Numabukuro       & & &  & \ensuremath{\bullet} & & & &42\\
25 & Araiyakushi-mae  & & &  & \ensuremath{\bullet} & & & &42\\
26 & Nakai            & &  & & \ensuremath{\bullet} & & & &42\\
27 & Shimo-Ochiai     & & &  & \ensuremath{\bullet} & & & &42\\
\midrule
\multicolumn{2}{l}{Total} & & & & & 2& 27& 18&42\\
\bottomrule
\end{tabular}

\vspace{0.5em}
\parbox{0.95\textwidth}{\footnotesize
Notes: \ensuremath{\bullet} indicates that the service stops at the station. Blank cells indicate that no trains of that service type stop at the station. The number of trains is based on trains arriving at the CBD (Seibu-Shinjuku) from the first train of the day through 10:30 a.m.
}

\end{table}

Crowding on the Seibu Shinjuku Line varies not only across trains but also between stations along each train's route. Across trains, those with different arrival times at the CBD and different service types carry different numbers of passengers, even during the morning peak. Figures~\ref{fig:app-players} and~\ref{fig:app-crowding} show the number of commuters on each train and the resulting crowding rates, respectively. Along its route, an inbound train in the morning typically becomes more crowded section by section as it approaches central Tokyo, because more passengers board than alight at each station. As a result, a commuter boarding near the suburban end experiences different crowding from one boarding closer to the CBD, even on the same train. We therefore measure crowding for each train and each section between stations, rather than for each train as a whole. Because each commuter's train choice adds to crowding on the sections they travel through, crowding levels are determined jointly by the choices of all commuters. Section~\ref{sec:model} models this interaction.

\section{Model}
\label{sec:model}
This section presents a model of commuters' train choices on a single rail line, based on the setting described in Section~\ref{sec:background}. We first describe the rail line, the commuters, and the trains available to them. We then specify the commuting cost of each train, which consists of schedule delay, travel time, and crowding costs. Finally, we define the equilibrium of the resulting population game and show that the game is a potential game, a property on which our estimation method in Section~\ref{sec:estimation} relies.

Consider a city with population $N$ consisting of $L$ zones and the CBD. Zones are labeled $1,2, ..., L$ in order of decreasing distance from the CBD, with the CBD indexed by $L+1$. Let $\tau_r^\ell$ denote the scheduled travel time of train $r$ over the physical rail segment from zone $\ell$ to zone $\ell+1$. Thus, segment travel times may differ across trains, including between local and express services. Commuters travel to the CBD by rail. Let $t^\ast$ denote the desired CBD arrival time, common across commuters in the baseline specification. It need not coincide with the work start time because commuters may need additional time to reach their workplace after arriving at the CBD. $R$ trains operate during the commuting period. Let $t_r$ denote the arrival time of train $r$ at the CBD. Trains are numbered in order of increasing arrival time at the CBD. Also, let $K_r$ denote the passenger capacity of train $r$.

Since the stopping stations differ by train type, let $\mathcal{R}^\ell \subseteq [R]$ denote the set of trains available to commuters living in zone $\ell$.\footnote{Given a natural number $x$, we define $[x]=\{1, 2, ..., x\}$.} In our application, each station on the line corresponds to a zone. We assume that each commuter rides a single train from their origin station to the CBD and do not model transfers between trains on the line, such as from a local train to an express train at an intermediate station. Also, let $E_r^k$ denote the number of exogenous passengers on train $r$ in segment $k$. Here, exogenous passengers are passengers on the line who are not commuters in the model, such as passengers traveling for purposes other than commuting and commuters who alight at an intermediate station before reaching the CBD (Section~\ref{subsec:ptsma}).

Let $s^\ell$ denote the share of commuters living in zone $\ell\in [L]$. The set of population states is given by $\mathcal{X}_N =\prod_{\ell=1}^L \mathcal{X}_N^\ell$ where
\begin{equation}
\mathcal{X}_N^\ell = \left\{ (x_r^\ell)_{r\in \mathcal{R}^\ell} \in \mathbb{R}_+^{|\mathcal{R}^\ell|}: \forall r\in \mathcal{R}^\ell, Nx_r^\ell \in \mathbb{N}, \sum_{r\in \mathcal{R}^\ell} x_r^\ell = s^\ell \right\}.
\end{equation}
$x_r^\ell$ denotes the share of people commuting from zone $\ell$ using train $r\in \mathcal{R}^\ell$, and we write a population state as $\bm x=(x_r^\ell)_{\ell\in[L],\, r\in\mathcal{R}^\ell}$. Here, we consider $N$ such that $Ns^\ell \in \mathbb{N}$ for all $\ell$. We also let $\mathcal{X}=\prod_{\ell=1}^L\mathcal{X}^\ell$ denote the continuous state space, where $\mathcal{X}^\ell$ is obtained from $\mathcal{X}_N^\ell$ by dropping the integrality constraint $Nx_r^\ell\in\mathbb{N}$.

Let
\begin{equation}
y_r^k(\bm x)=\sum_{j\in[k]:\, r\in\mathcal{R}^j} x_r^j
\end{equation}
denote the share of commuters on train $r$ in segment $k$. The in-vehicle crowding cost, i.e., the disutility of riding in a crowded car, when commuting from zone $\ell$ using train $r\in \mathcal{R}^\ell$ is
\begin{equation}
G_r^\ell (\bm x)= \sum_{k=\ell}^L \tau_r^k g\left(\frac{E_r^k+Ny_r^k(\bm x)}{K_r}\right)
\end{equation}
Since $E_r^k$ and $K_r$ are measured in passengers, $Ny_r^k(\bm x)$ is the number of endogenous commuters on train $r$ in segment $k$, and the argument of $g$ is the crowding rate of train $r$ in that segment. Here, $g(\cdot)$ represents the per-unit-time in-vehicle crowding cost as a function of the crowding rate, so that $\tau_r^k g(\cdot)$ measures crowding exposure over segment $k$.  If a train passes an intermediate station without stopping, its scheduled time between consecutive stops is allocated across the intervening physical segments in proportion to segment distance. Hence, the total allocated time equals the train's scheduled stop-to-stop travel time.  We assume the following linear function.
\begin{equation}
g(y)=\lambda y.
\end{equation}

Define the scheduled in-vehicle travel time from zone $\ell$ to the CBD as
\begin{equation}
T_r^\ell=\sum_{k=\ell}^L\tau_r^k.
\end{equation}
The commuting cost when commuting from zone $\ell$ using train $r\in \mathcal{R}^\ell$ is
\begin{equation}\label{eq:cost}
C_r^\ell (\bm x) = \delta T_r^\ell + G_r^\ell (\bm x)
 + \beta_E(t^\ast-t_r)_+ + \beta_L(t_r-t^\ast)_+,
\end{equation}
where $(a)_+=\max\{a,0\}$. The nonnegative parameters $\beta_E$ and $\beta_L$ measure the per-minute costs of early and late arrival, respectively, and may differ. The nonnegative parameter $\delta$ measures the cost of in-vehicle travel time, while $\lambda\geq0$ measures additional crowding disutility per unit of travel time and crowding rate. Thus, travel time affects cost both directly through $\delta T_r^\ell$ and through exposure to crowding in $G_r^\ell(\bm x)$. Costs common to all trains available at a given origin, such as the fare to the CBD (Section~\ref{sec:background}), do not affect train choice, since both the equilibrium condition and the choice probabilities below depend only on cost differences among trains in $\mathcal{R}^\ell$. We therefore normalize such costs to zero. The functions $G_r^\ell$ and $C_r^\ell$ are well defined for every $\bm x\in\mathcal{X}$. Since $\mathcal{X}_N\subset\mathcal{X}$, they also apply to finite-population states, and the extension to $\mathcal{X}$ is used in the large-population limit below.

Let $\bm e_r^\ell\in\prod_{m=1}^L\mathbb{R}^{|\mathcal{R}^m|}$ denote the vector whose $(\ell,r)$-entry is 1 and all other entries are 0. Then, a population state $\bar{\bm x} \in \mathcal{X}_N$ is a Nash equilibrium if
\begin{equation}
\forall \ell \in [L], \forall r, s\in \mathcal{R}^\ell, \bar{x}_r^\ell >0 \Rightarrow C_r^\ell(\bar{\bm x})\le C_s^\ell( \bar{\bm x}+ \tfrac{1}{N}(\bm e_s^\ell - \bm e_r^\ell)).
\end{equation}
That is, no commuter can reduce their commuting cost by unilaterally switching to another available train. The right-hand side evaluates the cost of train $s$ at the state after the switch, because the switching commuter adds to the crowding on train $s$.

This game is a potential game \citep{monderer1996GamesEcon.Behav.}: there exists a single function of the population state whose change under any unilateral switch equals the negative of the change in the switching commuter's cost. Since crowding costs depend only on the number of passengers on each train segment, such a function can be constructed as in congestion games \citep{rosenthal1973IntJGameTheory}. See also \citet[][Section 11.5]{sandholm2010} for finite-population potential games. Define
\begin{equation}\label{eq:finite-potential}
\begin{aligned}
\Phi_N(\bm x) ={}& -\sum_{\ell=1}^L\sum_{r\in\mathcal R^\ell} N x_r^\ell
\left\{\delta T_r^\ell+\beta_E(t^\ast-t_r)_+ +\beta_L(t_r-t^\ast)_+\right\}\\
&-\sum_{r=1}^R\sum_{k=1}^L \tau_r^k
\sum_{n=1}^{Ny_r^k(\bm x)} g\left(\frac{E_r^k+n}{K_r}\right).
\end{aligned}
\end{equation}
Then, for all $\ell\in[L]$, $r,s\in\mathcal R^\ell$, and $\bm x\in\mathcal X_N$ with $x_r^\ell>0$,
\begin{equation}\label{eq:potential-difference}
\Phi_N\!\left(\bm x+\tfrac1N(\bm e_s^\ell-\bm e_r^\ell)\right)-\Phi_N(\bm x)
= -\left[C_s^\ell\!\left(\bm x+\tfrac1N(\bm e_s^\ell-\bm e_r^\ell)\right)-C_r^\ell(\bm x)\right].
\end{equation}
Appendix~\ref{app:potential-game-condition} provides the derivation. That is, any unilateral switch that lowers a commuter's cost strictly increases $\Phi_N$. Hence every maximizer of $\Phi_N$ on the finite set $\mathcal X_N$ is a Nash equilibrium, and an equilibrium exists. The train-specific time and schedule-delay terms are independent of the population state and enter $\Phi_N$ linearly, so they do not affect the potential-game structure.

Consider the large-population limit $N\to\infty$ in which exogenous passengers and train capacities scale proportionally with the number of commuters, i.e., $E_r^k/N$ and $K_r/N$ are held fixed at their observed values. Then the crowding rate $(E_r^k+Ny_r^k(\bm x))/K_r$ depends only on $\bm x$, and the effect of a single commuter on crowding vanishes. In this limit, $\Phi_N/N$ converges uniformly to
\begin{equation}\label{eq:limit-potential}
\begin{aligned}
\phi(\bm x) ={}& -\sum_{\ell=1}^L\sum_{r\in\mathcal R^\ell}
\left\{\delta T_r^\ell+\beta_E(t^\ast-t_r)_+ +\beta_L(t_r-t^\ast)_+\right\}x_r^\ell\\
&-\sum_{r=1}^R\sum_{k=1}^L\tau_r^k\int_0^{y_r^k(\bm x)} g\left(\frac{E_r^k+Nz}{K_r}\right)dz,
\end{aligned}
\end{equation}
which is concave on $\mathcal X$ whenever $g$ is nondecreasing, because each integral is a convex function of a linear function of $\bm x$.

The potential is the key to our estimation. As we will see in Section~\ref{sec:estimation}, when commuters revise their train choices according to the logit choice rule, the potential-game structure gives the stationary distribution of the resulting process in closed form, with states of higher $\Phi_N$ being more likely. We use this distribution as the likelihood. Evaluating this likelihood directly is infeasible, because normalizing the distribution requires summing over all states in $\mathcal X_N$. In the large-population limit, however, the likelihood depends on $\phi$ and can be evaluated by maximizing a strictly concave function.

\section{Data}
\label{sec:data}

\subsection{The PTSMA data}
\label{subsec:ptsma}

We use individual-level microdata from the 2015 Public Transport Survey for Metropolitan Areas (PTSMA) in the Tokyo area.\footnote{The individual-level PTSMA data used in this study were provided by MLIT. All statistics derived from these data and reported in this paper were independently produced or processed by the authors and are not official statistics published by MLIT. The individual-level data are not publicly available, and access requires approval from MLIT.} The PTSMA is a large-scale passenger survey conducted to measure public transport use in Japan's three major metropolitan areas. The data contain detailed information on each respondent's railway travel on the survey day, as well as individual characteristics such as age and gender. For our analysis, we mainly use information on trip purpose, boarding and alighting stations, boarding and alighting times, work start time, and egress time, which is the travel time from the final railway station to the destination. The data also include expansion factors that convert the survey sample into the corresponding population. Section~\ref{subsec:expansion} describes how we use these expansion factors.

The PTSMA microdata have at least three advantages for our study. First, we can infer the train actually chosen by each respondent from their boarding and alighting stations and times. Although the data do not directly record the specific train used, we can match each respondent to a train on the Seibu Shinjuku Line by combining this information with the train timetable (Appendix~\ref{app:train-level-data}). To reflect service conditions at the time of the PTSMA survey, we take train schedules from the November 2015 issue of the \textit{Tokyo Timetable} \citep{Kotsu2015}. For trips involving multiple railway lines, the data record the boarding and alighting stations for each line used, but only the initial boarding time and the final alighting time for the entire railway trip. Thus, when the Seibu Shinjuku Line is the first or last line used, its boarding or alighting time is directly observed and can be used for train matching. When the Seibu Shinjuku Line is used between other railway lines, however, neither its boarding nor alighting time is directly observed. For these trips, we impute the boarding and alighting times on the Seibu Shinjuku Line by allocating the observed total travel time in proportion to the fare-calculation distance of each railway segment, and then match each respondent to a train.

Second, the data allow us to distinguish the players in our train-choice model from other passengers by jointly observing trip purpose and railway route. As defined in Section~\ref{sec:background}, the players in our model are commuters who use inbound trains on the Seibu Shinjuku Line and alight in the CBD, and each player's origin is the station at which they board the line, even if they transfer there from another line. By contrast, passengers who use the same inbound trains for non-commuting purposes, commuters who alight before reaching the CBD, and commuters without commuter passes (Section~\ref{subsec:expansion}) do not participate in the train-choice problem modeled in this study. These passengers nevertheless travel on the same trains as the players and therefore contribute to the onboard crowding experienced by them. We thus treat these passengers as exogenous passengers. Because the data jointly record trip purpose and railway route, we can construct both the endogenous commuter population and the exogenous passenger flow from a single survey, without combining separate data sources collected under different designs.

Third, because the data record both work start time and egress time, we can construct the desired CBD arrival time from information outside the train choices that our model seeks to explain, although this also requires an assumption about how early commuters wish to arrive at work (Section~\ref{subsec:desired-arrival}). This matters because the desired CBD arrival time determines schedule delay costs and therefore directly affects the estimated schedule delay preferences. Smart-card records, which are often used in empirical studies of rail crowding, generally lack such information, so studies using them infer desired arrival times from observed arrival times \citep[e.g.,][]{xu2024TransportationResearchPartB:Methodological}.

\subsection{Expansion factors}
\label{subsec:expansion}

The PTSMA microdata provide two sets of expansion factors that scale the survey sample up to the population of rail users (Appendix~\ref{app:ptsma-expansion}). Because the two sets are calibrated to different population benchmarks, we restrict the players to commuters who travel with commuter passes and use the factors based on commuter pass sales for them. We use these expansion factors to determine the size of each population in the model, but not the distribution of commuters across trains within a population. Specifically, the number of commuters boarding at each station is the sum of the expansion factors of the players who board there. This determines the population share \(s^\ell\) of each station and the total population \(N\). Within each station, we then allocate this number across trains in proportion to the unweighted numbers of survey respondents who choose each train. The observed share of commuters at station \(\ell\) who take train \(r\), \(x_r^\ell/s^\ell\), is therefore the share of survey respondents boarding at that station who take train \(r\). This approach preserves the station totals implied by the expansion factors while basing the train shares within each station directly on the observed choices of respondents. The expansion factors are determined by the ticket sales office at which each commuter pass was purchased and by sex and age, not by train choice. Within a station, however, they vary considerably across respondents, so train shares weighted by them can be dominated by a few respondents with large expansion factors, particularly at stations with few respondents. Unweighted shares avoid this source of variability. They would differ from the population shares only if train choices differed systematically across the groups that determine the expansion factors. The exogenous passenger counts \(E_r^k\) are also constructed from the expansion factors of the records of exogenous passengers (Appendix~\ref{app:train-level-data}). Because the expansion factors scale a survey sample up to the population, the \(N\) commuters in the model should not be interpreted as \(N\) independent observations. Appendix~\ref{app:survey-ess} discusses the implications for statistical inference.

\subsection{Desired CBD arrival time}
\label{subsec:desired-arrival}

We construct the desired CBD arrival time in two steps. First, we express each player's work start time relative to the time of arrival in the CBD, so that players who transfer to another railway line there and those who do not are treated alike. For players who transfer, we subtract the railway travel time after the CBD from the reported work start time. For players who do not transfer, we use the reported work start time as it is. We refer to this adjusted work start time as the CBD-adjusted work start time and denote it by \(w_i^{CBD}\) for player \(i\). Second, we define the desired CBD arrival time as
\begin{equation}
t_i^* = w_i^{CBD} - e_i - b.
\label{eq:desired-arrival}
\end{equation}
Here, \(e_i\) denotes egress time, and \(b\) denotes the preferred buffer time, defined as the number of minutes before the work start time that the commuter would like to arrive at the workplace. The preferred buffer time \(b\) reflects commuter preferences and is not directly observed in the PTSMA. We therefore use external survey evidence. Specifically, we use a 1999 survey conducted by Citizen Watch Co., Ltd., a Japanese watch manufacturer, among workers living in the Tokyo metropolitan area, which asked how many minutes before the scheduled work start time they would ideally like to arrive at work.\footnote{Citizen Watch Co., Ltd., ``Shihanseiki hikaku de miru bijinesuman/OL no seikatsu jikan ch\={o}sa'' [Daily time use of office workers: A quarter-century comparison], released June 10, 1999, \url{https://www.citizen.co.jp/research/19990610/06.html}.} Based on the weighted average of the reported responses, we set \(b=15\) minutes.

Although \eqref{eq:desired-arrival} defines a desired CBD arrival time for each player, we assume for simplicity in the baseline model that all players share a common desired CBD arrival time \(t^*\). We compute \(t^*\) by replacing \(w_i^{CBD}\) and \(e_i\) in the definition of \(t_i^*\) with their means across players, weighting each player by the average expansion factor of the players boarding at the same station, as in Section~\ref{subsec:expansion}. This gives \(t^*=8{:}10\) a.m. Figure~\ref{fig:app-desired-baseline} shows the distributions of CBD-adjusted work start times, egress times, and individual desired CBD arrival times. In Section~\ref{subsec:heterogeneity}, we relax this assumption and introduce heterogeneity in desired CBD arrival times by grouping players according to their CBD-adjusted work start times.

\subsection{Train schedules and capacities}
\label{subsec:timetable}

We obtain train arrival and departure times, travel times, and passenger capacities from the \textit{Tokyo Timetable} and the Railway Transport Service Survey in the PTSMA. For each train, we use the scheduled arrival and departure times at each station as reported in the \textit{Tokyo Timetable}, which records times at the minute level. We calculate travel time for each origin-destination pair as the difference between the departure time at the boarding station and the arrival time at the alighting station. We measure each train's capacity by its rated passenger capacity, which counts both seats and standing room. Passenger capacities come from the Railway Transport Service Survey, which reports capacity by railway line, direction, section between adjacent stations, and time period but not for individual trains. We therefore derive the capacity of each train by combining these period-level data with the timetable, which shows the trains operating in each period (Appendix~\ref{app:train-level-data}).

\subsection{Analysis sample}
\label{subsec:sample}

Our analysis sample consists of the players defined in Sections~\ref{subsec:ptsma} and~\ref{subsec:expansion}. We include trains that arrive in the CBD between the first train of the day and 10:30 a.m. This time window is wider than the conventional morning peak period so that we can capture commuters' crowding-avoidance behavior through shifts in CBD arrival time. Table~\ref{tab:sample-overview} reports descriptive statistics. The 2,740 respondents in the sample represent 140,473 commuters. The number of available trains ranges from 25 to 89 across stations, and about one-third of the station--train cells have no observed commuters, which we examine in Section~\ref{subsec:robustness}. Appendix~\ref{app:additional-descriptive} presents additional descriptive figures.

\begin{table}[htbp]
\centering
\caption{Descriptive statistics}
\label{tab:sample-overview}
\small
\begin{tabular}{lrrrr}
\toprule
\textbf{Panel A. Sample characteristics} & \textbf{Value} & & &\\
\midrule
Survey respondents (players) & 2,740 & & &\\
Expanded commuter population $N$ & 140,473 & & &\\
Origin stations & 27 & & &\\
Trains & 89 & & &\\
Available station--train cells & 1,251 & & &\\
Cells with a positive commuter count & 837 & & &\\
Unobserved cells & 414 (33.1\%) & & &\\
CBD arrival-time window & First train--10:30 & & &\\
\midrule
\textbf{Panel B. Commuter and train characteristics}
& \textbf{Mean} & \textbf{SD} & \textbf{Min} & \textbf{Max}\\
\midrule
Expanded commuters per boarding station & 5,202.7 & 5,072.9 & 651 & 26,058\\
Expanded commuters per train & 1,578.3 & 1,156.2 & 0 & 4,700\\
Available trains per boarding station & 46.3 & 17.0 & 25 & 89\\
Train capacity & 1,264.7 & 140.7 & 1,120 & 1,400\\
Travel time to CBD (min.) & 30.8 & 13.0 & 5 & 78\\
Expanded exogenous passengers per train-section & 342.5 & 372.7 & 0 & 1,918\\
\bottomrule
\end{tabular}

\vspace{0.5em}
\parbox{0.95\textwidth}{\footnotesize Notes: Survey respondents are the unweighted number of PTSMA respondents in the analysis sample. Commuter and exogenous passenger counts are expanded using the PTSMA expansion factors (Section~\ref{subsec:expansion}).}
\end{table}

\section{Estimation}
\label{sec:estimation}

This section describes how we estimate the cost parameters from the data described in Section~\ref{sec:data}. The data provide a single observed state $\bm x$, whose element $x_r^\ell$ is the share of commuters who board at station $\ell$ and take train $r$. We regard this state as a draw from the stationary distribution of a process in which commuters occasionally revise their train choices. Section~\ref{subsec:likelihood} derives the likelihood function from this distribution using the potential-game structure in Section~\ref{sec:model}. Section~\ref{subsec:model-based-bootstrap} describes the parametric bootstrap that we use to compute standard errors.

\subsection{Likelihood function}
\label{subsec:likelihood}

We formalize the revision process as a Markov chain. Let $\{X_N^t\}_{t\ge 0}$ be a Markov chain over $\mathcal{X}_N$. Each player receives a revision opportunity based on an independent Poisson process with a rate of 1. Collectively, revision opportunities for players follow a Poisson process with rate $N$, meaning that the expected duration for which the population state remains unchanged is $1/N$. 

When a player receives a revision opportunity, the player updates their train choice according to the {\it logit protocol} with noise level $\eta > 0$ \citep[][Section 11.5.2]{sandholm2010}. The protocol operates as follows: Suppose a player in zone $\ell\in [L]$ currently taking train $r\in \mathcal{R}^\ell$ receives a revision opportunity. Then, the player switches to train $s\in \mathcal{R}^\ell$ with the following probability:
\begin{equation}
    \frac{\exp[- \eta^{-1}C_s^\ell(\bm x+\frac{1}{N}(\bm e_s^\ell -\bm e_r^\ell))]}{\sum_{q\in \mathcal{R}^\ell} \exp[- \eta^{-1}C_q^\ell(\bm x+\frac{1}{N}(\bm e_q^\ell -\bm e_r^\ell))]}.
\end{equation}

Let $\varsigma_0=0$ and let $\varsigma_i$ denote the random time at which the $i$-th revision opportunity occurs, and define $Y_i=X_N^{\varsigma_i}$ for $i=0,1,2,\dots$. The probability that the revising player lives in zone $\ell$ and currently takes train $r$ is $x_r^\ell$. Since the revising player is drawn uniformly and the protocol does not depend on time, $\{Y_i\}_{i\ge 0}$ is a time-homogeneous Markov chain: for $r,s\in \mathcal{R}^\ell$,
\begin{equation}\label{eq:Markov}
    \Pr\left[ Y_{i+1} = \bm x + \frac{1}{N} (\bm e_s^\ell - \bm e_r^\ell) \Big| Y_i= \bm x \right] 
    = 
   x_r^\ell    \frac{\exp[- \eta^{-1}C_s^\ell(\bm x+\frac{1}{N}(\bm e_s^\ell -\bm e_r^\ell))]}{\sum_{q\in \mathcal{R}^\ell} \exp[- \eta^{-1}C_q^\ell(\bm x+\frac{1}{N}(\bm e_q^\ell -\bm e_r^\ell))]}.
\end{equation}
which does not depend on $i$. Because revision opportunities arrive at the constant rate $N$ regardless of the state, $\{X_N^t\}_{t\ge0}$ and $\{Y_i\}_{i\ge0}$ have the same stationary distribution. According to Theorem 11.5.12 of \citet{sandholm2010}, it is uniquely given by
\begin{equation}\label{eq:stationary_dist}
    \mu_N(\bm x) = \frac{1}{\kappa_N} \prod_{\ell\in [L]} \frac{(Ns^\ell)!}{\prod_{r\in \mathcal{R}^\ell}(Nx_r^\ell)!} \exp [ \eta^{-1}\Phi_N(\bm x)],
\end{equation}
for all $\bm x\in \mathcal{X}_N$, where $\kappa_N$ is determined so that $\sum_{\bm x \in \mathcal{X}_N} \mu_N(\bm x)=1$, and $\mu_N$ denotes the probability distribution over states. Thus, it can be used as a likelihood function. 

However, a computational difficulty arises due to the normalizing constant $\kappa_N$, because the probabilities over all possible states should be computed. This is well recognized in the literature on the structural estimation of network games. Among others, \citet{mele2017Econometrica} adopted the Bayesian approach, where an advanced version of the Metropolis--Hastings algorithm was used to bypass the computation of $\kappa_N$.

In our context of commuting, the total number of players can be considered large. We then consider the fact that $\frac{1}{N}\ln \mu_N$ is uniformly convergent in $N$, and the limit function does not include the normalizing constant. Specifically, according to Theorem 12.2.3 of \citet{sandholm2010},
\begin{equation}
    \lim_{N\to\infty} \max_{\bm x\in \mathcal{X}_N}\left| \frac{1}{N} \ln \mu_N(\bm x) - \left\{ \Phi^\ast (\bm  x)- \max_{\bm  x'\in \mathcal{X}} \Phi^\ast(\bm x')\right\}\right| =0,
\end{equation}
where
\begin{equation}
\Phi^\ast(\bm x) = \frac{1}{\eta}\phi(\bm x)
-\sum_{\ell\in[L]}\sum_{r\in\mathcal R^\ell}x_r^\ell\ln x_r^\ell,
\end{equation}
and $\phi$ is the limit potential defined in \eqref{eq:limit-potential}.
Because the commuting cost \eqref{eq:cost} is linear in $(\beta_E,\beta_L,\lambda,\delta)$ under the linear specification of $g$, the likelihood depends on these parameters and $\eta$ only through their ratios. We therefore estimate $\theta=(\beta_E/\eta,\beta_L/\eta,\lambda/\eta,\delta/\eta)$ under nonnegativity constraints, treating $t^\ast$ as fixed in each specification.

Therefore, under the supposition that $N$ is sufficiently large, the log likelihood $\ln \mu_N(\bm x)$ is proportional to $\Phi^\ast (\bm x)- \max_{\bm x'\in \mathcal{X}} \Phi^\ast(\bm x')$, which replaces the task of computing the normalizing constant with the task of maximizing the potential function. In general, maximizing a function is not an easy task. However, when $g$ is nondecreasing, $\phi$ is concave on $\mathcal X$, as shown in Section~\ref{sec:model}. Hence, $\Phi^\ast$ is strictly concave because of the entropy term $-\sum_{\ell}\sum_{r}x_r^\ell\ln x_r^\ell$.

Our log-likelihood function is then given by
\begin{equation}
   \ell (\theta | \bm x) = \Phi^\ast (\bm x) - \max_{\bm x' \in \mathcal{X}}\Phi^\ast(\bm x').
   \label{eq:log-likelihood_function}
\end{equation}
Then we search for the maximum likelihood estimator $\hat{\theta}\in \arg\max_\theta \, \ell(\theta |\bm x)$. Since $\Phi^\ast$ is strictly concave in $\bm x$, the inner maximization over $\mathcal X$ has a unique solution and can be solved reliably. Moreover, because $\Phi^\ast$ is linear in $\theta$ for each $\bm x$, $\max_{\bm x'\in\mathcal X}\Phi^\ast(\bm x';\theta)$ is convex in $\theta$ as the pointwise maximum of affine functions \citep[][Section 3.2.3]{boyd2004}, and hence the log-likelihood $\ell(\theta\mid\bm x)$ is concave in $\theta$.
Appendix~\ref{app:numerical-optimization} describes the nested numerical
optimization and the associated convergence checks.

\subsection{Parametric bootstrap}
\label{subsec:model-based-bootstrap}

To estimate the standard errors of $\widehat\theta$, we use a parametric
bootstrap based on the finite-population equilibrium distribution.  Let
$\widehat\mu_N$ denote the stationary distribution in
\eqref{eq:stationary_dist} evaluated at $\widehat\theta$.  We treat
$\widehat\mu_N$ as the data-generating process: we draw population states
$\bm x^1,\ldots,\bm x^B$ from $\widehat\mu_N$, maximize the same likelihood
as in \eqref{eq:log-likelihood_function} for each $\bm x^b$ to obtain
$\widehat\theta^{b}$, and measure the sampling variability of
$\widehat\theta$ by the dispersion of $\widehat\theta^{1},\ldots,\widehat\theta^{B}$.
The draws are obtained by simulating a Markov chain, as described below.

Direct sampling from $\widehat\mu_N$ is difficult because its normalizing
constant is unavailable.  We therefore simulate the logit revision process
in \eqref{eq:Markov}, for which $\widehat\mu_N$ is the invariant
distribution.  As described in Section~\ref{subsec:likelihood}, in each
period of this process one commuter receives a revision opportunity and
may switch only among the trains available at their origin station.  The
total population and every station-specific population are therefore held
fixed throughout the simulation.

Sampling from a Markov chain requires choosing two settings: a burn-in,
the number of initial periods discarded so that the chain approaches
$\widehat\mu_N$ from its initial state, and a sampling interval, the number
of periods between retained states, which reduces their dependence.  We
measure both in units of $N$ periods.  Because one commuter is selected
uniformly for a revision opportunity in each period, $N$ periods correspond
to one revision opportunity per commuter on average.  We initialize the
chain at the observed allocation, discard the first $10N$
periods, and then
retain a population state every $N$ periods, for a total of $B=1{,}000$
retained states.\footnote{If the initial state were a draw from
$\widehat\mu_N$, the chain would be stationary from the start and no
burn-in would be needed.  The observed allocation, however, is modeled as a
realization from the equilibrium distribution at the true parameter rather
than at $\widehat\theta$, and $\widehat\theta$ is itself estimated from
this allocation.  The observed allocation therefore need not be a typical
state under $\widehat\mu_N$, and the early periods of the chain may reflect
the transition away from it rather than draws from $\widehat\mu_N$.
Appendix~\ref{app:bootstrap-diagnostics} shows that introducing the $10N$
burn-in reduces the bootstrap standard errors, whereas increasing it further
to $20N$ has almost no effect.}  In terms of the chain $\{Y_i\}_{i\ge0}$ in
Section~\ref{subsec:likelihood}, simulated at $\widehat\theta$, the retained
states are
\begin{equation}
    \bm x^b=Y_{(10+b)N},
    \qquad b=1,\ldots,B,
\end{equation}
and the bootstrap standard error is the sample standard deviation of the
corresponding estimates:
\begin{equation}
    \widehat{\operatorname{SE}}_{\mathrm{boot}}
    =
    \left\{
    \frac{1}{B-1}\sum_{b=1}^{B}
    \left(
      \widehat\theta^{b}-\overline{\widehat\theta}
    \right)^2
    \right\}^{1/2},
    \qquad
    \overline{\widehat\theta}
    =\frac{1}{B}\sum_{b=1}^{B}\widehat\theta^{b}.
    \label{eq:bootstrap-se}
\end{equation}

Because the retained states need not be independent, we assess the
sensitivity of the bootstrap mean, bias, standard error, and percentile
interval to the number of retained states, burn-in, and sampling interval.
We also quantify the finite-simulation error of each bootstrap standard
error by its Monte Carlo standard error (MCSE), computed from the
squared-centered series, and use a relative MCSE of 5\% as our operational
precision target.  This target corresponds to about 200 effectively
independent replications, the number that \citet[Section~6.4]{efron1993}
suggest is rarely exceeded for estimating a standard error.  Appendix~\ref{app:finite-simulation-error} gives the
details.

The parametric bootstrap conditions on the expanded commuter population
and therefore does not account for uncertainty from survey sampling or
from the construction of the expansion weights
(Appendix~\ref{app:expansion-weight-inference}).
Section~\ref{subsec:robustness} examines the sensitivity of the estimates to
this uncertainty by perturbing the observed train allocations at the stations
with the fewest respondents.

\section{Results}
\label{sec:results}

We first report the baseline estimates, then allow desired CBD arrival times to differ across commuters, and finally report robustness checks on the simulation settings for the bootstrap, expansion weights and stations with few survey respondents, and the treatment of unobserved cells.

\subsection{Baseline results}
\label{subsec:baseline}

Table~\ref{tab:baseline-results} reports the baseline estimates with the desired CBD arrival time, $t^*$, set to 8:10, together with estimates using alternative values of 8:00 and 8:20. Columns 1, 2, and 3 report the results for $t^*=8{:}00$, 8:10, and 8:20, respectively. For each specification, the table reports the estimated cost parameters and the standard errors and 95\% confidence intervals based on the bootstrap. Table~\ref{tab:bootstrap-results} provides additional details on the bootstrap results.

\begin{table}[htbp]
\centering
\caption{Baseline estimation results}
\label{tab:baseline-results}
\small

\begin{tabular}{lrrr}
\toprule
& \multicolumn{3}{c}{Desired CBD arrival time $t^\ast$} \\
\addlinespace
Parameter & 8:00 & 8:10 (Baseline) & 8:20 \\
\midrule
$\beta_E/\eta$   & 0.020949 & 0.018042 & 0.014786 \\
                  & (0.000148) & (0.000128) & (0.000108) \\
                  & [0.020643, 0.021206] & [0.017787, 0.018266] & [0.014567, 0.014979] \\
\addlinespace
$\beta_L/\eta$   & 0.019392 & 0.022468 & 0.024944 \\
                  & (0.000122) & (0.000136) & (0.000154) \\
                  & [0.019127, 0.019608] & [0.022172, 0.022707] & [0.024627, 0.025215] \\
\addlinespace
$\lambda/\eta$   & 0.001939 & 0.001845 & 0.000985 \\
                  & (0.000168) & (0.000166) & (0.000160) \\
                  & [0.001564, 0.002224] & [0.001480, 0.002169] & [0.000645, 0.001277] \\
\addlinespace
$\delta/\eta$    & 0.044960 & 0.046225 & 0.046448 \\
                  & (0.000681) & (0.000647) & (0.000668) \\
                  & [0.043695, 0.046360] & [0.045028, 0.047521] & [0.045262, 0.047842] \\
\bottomrule
\end{tabular}

\vspace{0.5em}
\parbox{0.95\textwidth}{\footnotesize Notes: $\beta_E/\eta$ and $\beta_L/\eta$ denote the cost parameters for early and late arrival, respectively, $\lambda/\eta$ denotes the crowding parameter, and $\delta/\eta$ denotes the travel-time parameter. Point estimates are reported in the first row for each parameter, bootstrap standard errors in parentheses, and 95\% percentile intervals in brackets. The baseline desired CBD arrival time is 8:10. All parameters are estimated subject to nonnegativity constraints. The bootstrap uses $B=1{,}000$ retained estimates, a sampling interval of $N$, and a burn-in of $10N$. Both the inner potential maximization and the outer likelihood maximization terminated successfully.}
\end{table}

Across all three values of $t^*$, the estimated cost parameters for
schedule delay early, schedule delay late, crowding, and travel time
are positive, and their 95\% confidence intervals exclude zero.
The schedule delay early parameter decreases as $t^*$ is shifted later,
whereas the schedule delay late parameter increases. This pattern reflects how the model rationalizes a given distribution of arrival times. When $t^*$ is shifted later, more of the observed arrivals occur before $t^*$ and fewer occur after it. The model attributes the larger share of early arrivals to a lower cost of arriving early and the smaller share of late arrivals to a higher cost of arriving late.\footnote{The same relationship appears in the bottleneck model \citep{vickrey1969Am.Econ.Rev.,arnott1990JournalofUrbanEconomics}, in which the equilibrium share of commuters who arrive early equals $\beta_L/(\beta_E+\beta_L)$.} The travel-time
parameter is relatively stable across the three specifications, ranging
from 0.044960 at $t^*=8{:}00$ to 0.046448 at $t^*=8{:}20$.
By contrast, the crowding parameter is similar at $t^*=8{:}00$ and
$t^*=8{:}10$, at 0.001939 and 0.001845, respectively, but falls to
0.000985 at $t^*=8{:}20$.

\begin{figure}[htbp]
    \centering
    \includegraphics[width=0.85\textwidth]{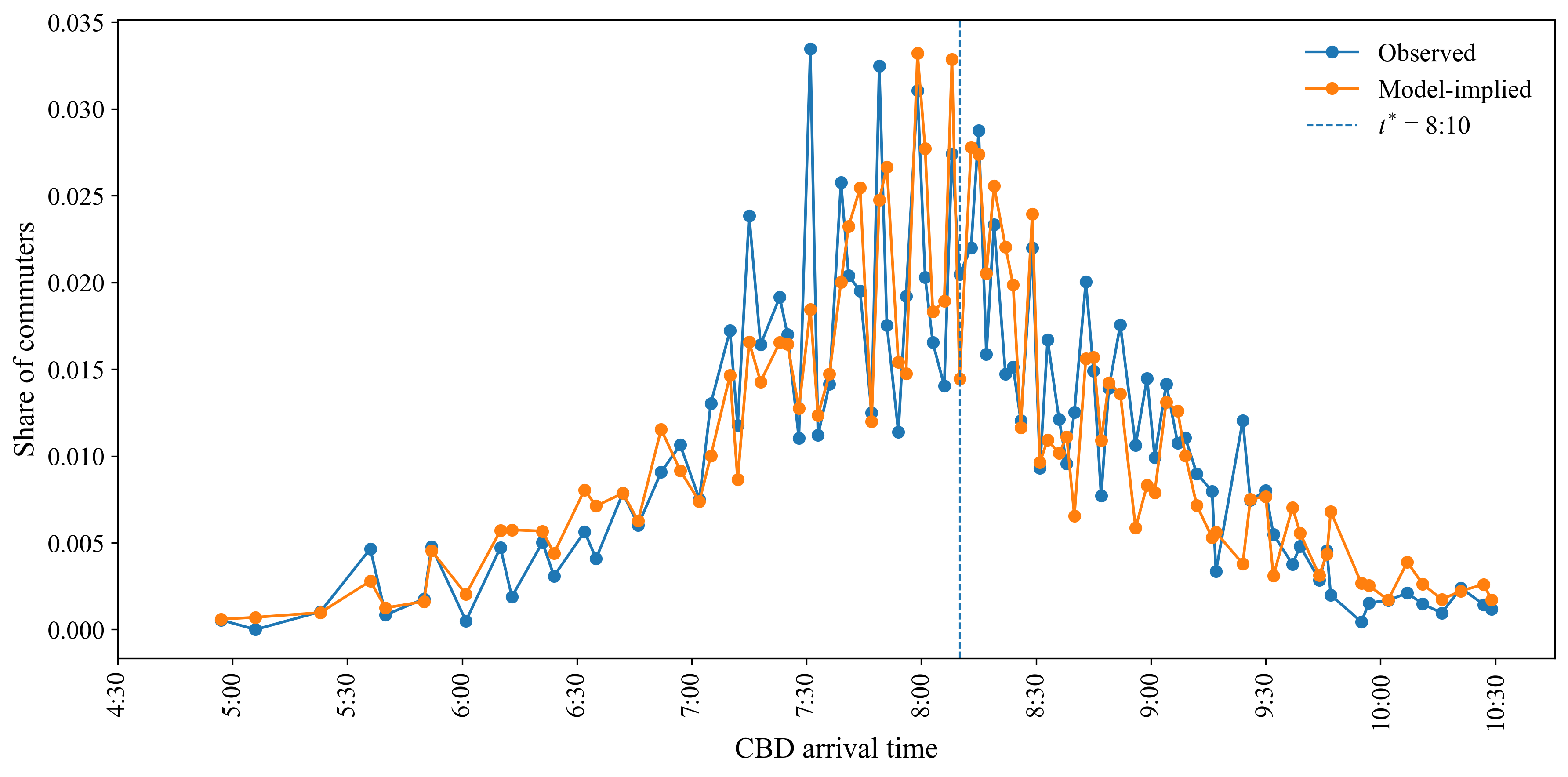}
    \caption{Observed and model-implied CBD arrival distributions}
    \label{fig:baseline-reproduction}

    \vspace{0.5em}
    \parbox{0.90\textwidth}{\footnotesize Notes: The figure compares the observed distribution of players' CBD arrival times with the model-implied distribution under the baseline specification with $t^*=8{:}10$. Both distributions are measured at the train level. The horizontal axis shows the CBD arrival time of each train at Seibu-Shinjuku Station.}
\end{figure}

Figure~\ref{fig:baseline-reproduction} compares the observed distribution of players' CBD arrival times with the model-implied distribution under the baseline specification with $t^*=8{:}10$. Although some differences remain in train-level fluctuations, the model broadly reproduces the overall shape of the observed distribution, including the increase toward the commuting peak and the subsequent decline.

To compare the cost components, we express each in travel-time equivalents by dividing it by the travel-time parameter. Under the baseline specification with $t^*=8{:}10$, one minute of schedule delay early and late is equivalent to 0.39 and 0.49 minutes of in-vehicle travel time, respectively. One minute of travel at a crowding rate of 100\% adds a crowding cost equivalent to 0.04 minutes. Across the three values of $t^*$, these equivalents range from 0.32 to 0.47 minutes for schedule delay early, from 0.43 to 0.54 minutes for schedule delay late, and from 0.021 to 0.043 minutes for crowding at a rate of 100\%. Crowding costs are therefore smaller than schedule delay costs in all three specifications.

\subsection{Heterogeneity in desired CBD arrival time}
\label{subsec:heterogeneity}

This subsection allows the desired CBD arrival time, $t^*$, to differ across commuters, whereas the baseline specification assumes a common $t^*$. We group commuters in two ways and assign a separate $t^*$ to each group. The first groups commuters by transfer status, distinguishing those who complete their trips at Takadanobaba or Seibu-Shinjuku from those who transfer there to another railway line. The second groups commuters more finely by their CBD-adjusted work start times.

\subsubsection{Heterogeneity by transfer status}
\label{subsubsec:heterogeneity-transfer}

We divide commuters into two groups according to their transfer status. We refer to commuters whose rail trips end at the CBD as terminating commuters and to those who transfer there to another railway line as transferring commuters. We set a separate desired CBD arrival time, $t^*$, for each group. The weighted average CBD-adjusted work start time is 9:02 for terminating commuters and 8:35 for transferring commuters, while the corresponding weighted average egress times are 15.8 and 10.1 minutes, respectively. Figures~\ref{fig:app-work-start-transfer} and~\ref{fig:app-egress-transfer} show the distributions of these work start times and egress times by group. Using the same preferred buffer time of 15 minutes as in the baseline specification, we set $t^*=8{:}30$ for terminating commuters and $t^*=8{:}10$ for transferring commuters. Figure~\ref{fig:app-desired-transfer} shows the distributions of individual desired CBD arrival times by group.

\begin{table}[htbp]
\centering
\caption{Heterogeneity by transfer status}
\label{tab:heterogeneity-transfer}
\small
\begin{tabular}{lrrr}
\toprule
Parameter & Estimate & Bootstrap SE & 95\% interval \\
\midrule
$\beta_E/\eta$ & 0.017300 & 0.000115 & [0.017067, 0.017511] \\
$\beta_L/\eta$ & 0.024218 & 0.000141 & [0.023921, 0.024485] \\
$\lambda/\eta$ & 0.001525 & 0.000157 & [0.001202, 0.001813] \\
$\delta/\eta$  & 0.046622 & 0.000709 & [0.045299, 0.048097] \\
\bottomrule
\end{tabular}

\vspace{0.5em}
\parbox{0.95\textwidth}{\footnotesize Notes: Desired CBD arrival time is set to 8:30 for commuters whose rail trips terminate on the Seibu Shinjuku Line and to 8:10 for commuters who transfer to another railway line at the CBD. The bootstrap SE is the sample standard deviation of the 1,000 retained estimates. The 95\% intervals are the 2.5th and 97.5th percentiles of those estimates.}
\end{table}

Table~\ref{tab:heterogeneity-transfer} reports the estimated cost parameters, together with bootstrap standard errors and 95\% confidence intervals (see Table~\ref{tab:bootstrap-transfer} for detailed bootstrap results). All four parameters are positive, and their 95\% confidence intervals exclude zero. In travel-time equivalents, one minute of schedule delay early and late corresponds to 0.37 and 0.52 minutes of in-vehicle travel time, respectively, close to the baseline values of 0.39 and 0.49. The additional crowding cost at a crowding rate of 100\% corresponds to 0.033 minutes, somewhat smaller than the baseline value of 0.04. Allowing desired CBD arrival times to differ by transfer status thus leaves the relative magnitudes of the cost components largely unchanged.

\subsubsection{Heterogeneity in CBD-adjusted work start times}
\label{subsubsec:heterogeneity-workstart}

We next allow the desired CBD arrival time, $t^*$, to vary with each commuter's CBD-adjusted work start time. For each commuter, we calculate $t^*$ by subtracting a common egress time of 11.2 minutes and a preferred buffer time of 15 minutes from the CBD-adjusted work start time. We use a common egress time rather than individual egress times so as not to lose players whose egress time is missing. For estimation, we group the resulting values of $t^*$ into 30- or 60-minute intervals labeled by their midpoints, so that the 8:00 group covers 7:45--8:14 with 30-minute intervals and 7:30--8:29 with 60-minute intervals. For each width, we consider two placements of the midpoints that differ by half the interval width, such as 8:00 and 8:15 for 30-minute intervals. This yields four specifications.

\begin{table}[htbp]
\centering
\caption{Heterogeneity in CBD-adjusted work start times}
\label{tab:heterogeneity-workstart}
\small

\begin{tabular}{lrrrr}
\toprule
& \multicolumn{2}{c}{30-minute grouping} & \multicolumn{2}{c}{60-minute grouping} \\
\cmidrule(lr){2-3}\cmidrule(lr){4-5}
Parameter & 8:00, 8:30 & 8:15, 8:45 & 8:00, 9:00 & 8:30, 9:30 \\
\midrule
$\beta_E/\eta$ & 0.023595 & 0.025518 & 0.021836 & 0.021861 \\
& (0.000094) & (0.000096) & (0.000089) & (0.000089) \\
& [0.023440, 0.023818] & [0.025350, 0.025727] & [0.021673, 0.022032] & [0.021710, 0.022061] \\
\addlinespace
$\beta_L/\eta$ & 0.076791 & 0.077389 & 0.056393 & 0.058421 \\
& (0.000365) & (0.000358) & (0.000257) & (0.000254) \\
& [0.076198, 0.077608] & [0.076694, 0.078127] & [0.055974, 0.056960] & [0.057939, 0.058970] \\
\addlinespace
$\lambda/\eta$ & 0.000000 & 0.001368 & 0.000000 & 0.000000 \\
& (0.000081) & (0.000153) & (0.000070) & (0.000073) \\
& [0.000000, 0.000280] & [0.001037, 0.001619] & [0.000000, 0.000230] & [0.000000, 0.000243] \\
\addlinespace
$\delta/\eta$ & 0.048824 & 0.052377 & 0.044211 & 0.046567 \\
& (0.000615) & (0.000675) & (0.000548) & (0.000563) \\
& [0.047478, 0.049902] & [0.051267, 0.053825] & [0.043064, 0.045198] & [0.045339, 0.047578] \\
\bottomrule
\end{tabular}

\vspace{0.5em}
\parbox{0.95\textwidth}{\footnotesize Notes: Commuters are grouped by $t^*$ derived from CBD-adjusted work start times, in 30-minute or 60-minute intervals. Column headings show the group midpoints, either on the hour and half hour or at a quarter past and a quarter to the hour for the 30-minute grouping, and either on the hour or on the half hour for the 60-minute grouping. Point estimates are reported in the first row for each parameter, bootstrap standard errors in parentheses, and 95\% percentile intervals in brackets.}
\end{table}

Table~\ref{tab:heterogeneity-workstart} reports the results (see Tables~\ref{tab:bootstrap-30min} and~\ref{tab:bootstrap-60min} for detailed bootstrap results). The parameters for schedule delay early, schedule delay late, and travel time are positive under all four groupings, and their 95\% confidence intervals exclude zero. The crowding parameter, by contrast, is sensitive to the grouping of $t^*$. Its confidence interval excludes zero only under the 30-minute grouping with midpoints at a quarter past and a quarter to the hour, where it is 0.001368. Under the other groupings, it is zero or close to zero, and the lower bound of its confidence interval is zero.\footnote{The small crowding parameter might reflect misspecification of schedule delay costs, because arriving late at work may entail a fixed penalty in addition to a cost proportional to the minutes of delay \citep{small1982Am.Econ.Rev.}. We therefore also estimate a specification with a late-arrival dummy that assigns a fixed cost when a commuter arrives after $t^*$. The dummy is positive under all four groupings, and its confidence intervals exclude zero. Including it lowers the schedule delay late parameter, but the crowding parameter remains zero or close to zero except under the 30-minute grouping with midpoints at a quarter past and a quarter to the hour (Tables~\ref{tab:bootstrap-30min} and~\ref{tab:bootstrap-60min}).}

In travel-time equivalents, one minute of schedule delay early corresponds to 0.47--0.49 minutes of in-vehicle travel time across these groupings, and one minute of schedule delay late corresponds to 1.25--1.57 minutes, compared with 0.39 and 0.49 minutes in the baseline. The additional crowding cost at a crowding rate of 100\% ranges from zero to 0.026 minutes. Allowing for heterogeneity in CBD-adjusted work start times thus raises the valuation of schedule delay, particularly of late arrival, while the valuation of crowding remains small.

\subsection{Robustness checks}
\label{subsec:robustness}

We conduct three robustness checks, which concern the simulation settings of the parametric bootstrap, expansion weights and stations with few survey respondents, and the treatment of unobserved station--train cells. Appendices~\ref{app:bootstrap-diagnostics}, \ref{app:survey-ess}, and \ref{app:unobserved-cells} report the details.

First, we examine the sensitivity of the parametric bootstrap results to simulation settings. Specifically, the bootstrap means, standard errors, and 95\% confidence intervals change little when we vary the number of retained states, sampling interval, and burn-in (Table~\ref{tab:bootstrap-diagnostics}). We also calculate the Monte Carlo standard errors (MCSEs) of the bootstrap standard errors, accounting for serial correlation in the Markov chain. The relative MCSEs range from 2.48\% to 2.78\%, below the 5\% precision target set in Section~\ref{subsec:model-based-bootstrap} (Table~\ref{tab:bootstrap-mcse}). In addition, the bootstrap means and standard errors remain stable across three different random seeds (Table~\ref{tab:bootstrap-seed-robustness}). These results indicate that the bootstrap results are not highly sensitive to simulation settings and that finite-simulation errors are relatively small.

Second, we examine two issues related to expansion weights and survey sample sizes. The first concerns statistical inference. The parametric bootstrap used in this study treats the expanded population as given and therefore does not account for uncertainty arising from survey sampling. Because the train shares within each station are based on unweighted respondent counts, their precision at each station depends on its number of respondents. Weighting the shares by the expansion factors instead would reduce the effective sample size, which measures the precision of weighted shares in terms of an equivalent number of unweighted respondents, to 62.0\% of the 2,740 respondents in total (Table~\ref{tab:survey-ess-by-station}). The unweighted shares are therefore more precise. The second concerns stations with few respondents, where sampling uncertainty in the train shares is largest. Because the bootstrap does not capture this uncertainty, we re-estimate the model after changing the allocation to the train with the largest number of commuters by $\pm$10\% and $\pm$20\% at Hon-Kawagoe and Nakai, the two stations with the fewest respondents. The estimated parameters change little across these specifications (Table~\ref{tab:low-ess-sensitivity}). These results indicate that the estimates are not highly sensitive to observed train allocations at stations with few respondents.

Third, we examine the robustness of the estimates to the treatment of unobserved station--train cells. In this study, we treat station--train cells with no observed commuters as having zero commuters, even when the timetable shows that the train stops at the station. Of the 1,251 available station--train cells, 414 cells (33.1\%) have no observed commuters (Table~\ref{tab:unobserved-cells-by-station}). To assess the sensitivity of the estimates to this treatment, we combine adjacent stations with the same choice set and re-estimate the model using 24-zone, 19-zone, and 15-zone specifications, in addition to the 27-station baseline. The share of unobserved cells decreases from 33.1\% in the baseline to 23.2\% in the 15-zone specification (Table~\ref{tab:station-aggregation-statistics}). The estimated parameters change by less than 1\% for schedule delay early, schedule delay late, and travel time, while the maximum change in the crowding parameter is 6.09\% (Table~\ref{tab:unobserved-cell-sensitivity}). The estimates thus remain largely stable as station aggregation reduces the share of unobserved cells, which suggests that treating these cells as zeros does not drive the baseline results.

\section{Discussion}
\label{sec:discussion}

This section places our estimates in the context of previous studies and discusses their policy implications. Our baseline estimates indicate that commuters do have a preference for avoiding crowding, but that the marginal disutility of crowding is smaller than that of schedule delay (Section~\ref{subsec:baseline}).\footnote{This comparison concerns only the relative magnitude of marginal disutility for an individual commuter and should not be interpreted as implying that crowding is unimportant for train choice or social welfare. Crowding disutility accumulates over the duration of the trip, and a large number of commuters may be exposed to the same crowded conditions. Therefore, the aggregate welfare loss from crowding cannot be inferred from the magnitude of individual marginal disutility alone and may still be substantial.} We first compare these estimates with external benchmarks and then discuss what they imply for policies to reduce peak-hour crowding.

Table~\ref{tab:external-benchmarks} summarizes the external benchmarks in travel-time equivalents. For crowding, the first benchmark is the MLIT \textit{Evaluation Manual for Railway Projects}, which is used to appraise railway projects in Japan and adopts a piecewise-linear crowding-disutility function whose slope increases with the crowding rate. Panel A of the table compares our crowding costs with the values implied by this function, which measures crowding in the same way as we do. Our crowding costs are similar to the MLIT values at crowding rates of 100\% and 150\% but smaller at 200\%. The other studies measure crowding differently. \citet{wardman2011TransportReviews}, a meta-analysis of British stated-preference studies, report multipliers that imply additional costs of 0.05--0.55 minutes for seated travel and 0.62--1.44 minutes for standing travel at seating load factors of 100--200\%. \citet{xu2024TransportationResearchPartB:Methodological}, who jointly model travel time, schedule delay, and crowding for commuters on the Beijing Subway Batong Line, report additional costs of 0.27--0.54 minutes at standee densities of 3--6 passengers per square meter. These values are larger than ours. The difference in crowding measures is unlikely to explain this gap. Because Japanese rated capacity includes standing room, a crowding rate of 100\% corresponds to a higher passenger density than a seating load factor of 100\%, at which all seats are just occupied. Expressing the benchmarks in terms of our measure would therefore widen, rather than narrow, the gap. One possible reason for the gap is that our baseline specification assumes that crowding disutility is linear in the crowding rate, an issue to which we return in Section~\ref{sec:conclusion}.

\begin{table}[!htbp]
\centering
\caption{Comparison with external estimates}
\label{tab:external-benchmarks}
\small
\begin{tabular}{lccc}
\toprule
\multicolumn{4}{l}{\textit{Panel A: Additional crowding cost per minute of travel}} \\
\addlinespace
 & \multicolumn{3}{c}{Crowding rate} \\
\cmidrule(lr){2-4}
Source & 100\% & 150\% & 200\% \\
\midrule
This study & 0.040 & 0.060 & 0.080 \\
MLIT appraisal manual & 0.027 & 0.069 & 0.160 \\
\addlinespace
\multicolumn{4}{l}{\textit{Panel B: Cost per minute of schedule delay}} \\
\addlinespace
Source & Setting and data & Early & Late \\
\midrule
This study & Tokyo, survey & 0.39 & 0.49 \\
\citet{kobayashi2020J.JSCESer.D3} & Tokyo, RP & 0.36 & 0.77 \\
\citet{peer2016TransportationResearchPartB:Methodological} & Netherlands, RP & 0.43 & 0.36 \\
\citet{xu2024TransportationResearchPartB:Methodological} & Beijing, smart card & 1.65 & 1.86 \\
\bottomrule
\end{tabular}

\vspace{0.5em}
\parbox{0.95\textwidth}{\footnotesize Notes: All values are in minutes of in-vehicle travel time. Panel A reports the additional cost of one minute of travel at each crowding rate, defined relative to Japanese rated capacity, which includes standing room. The MLIT values are based on the piecewise-linear crowding-disutility function in the MLIT \textit{Evaluation Manual for Railway Projects}. Panel B reports the costs of one minute of schedule delay early and late. For the other studies, these are the ratios of the schedule-delay coefficients to the travel-time coefficient, namely $\beta/\alpha$ and $\gamma/\alpha$ in \citet{kobayashi2020J.JSCESer.D3}, the MNL values for the morning commute of 6.62 and 5.56 euros per hour divided by 15.5 euros per hour in \citet{peer2016TransportationResearchPartB:Methodological}, and $\tilde\beta/\tilde\alpha$ and $\tilde\gamma/\tilde\alpha$ in the multi-class inverse-optimization estimates of \citet{xu2024TransportationResearchPartB:Methodological}. RP denotes revealed-preference data.}
\end{table}

Panel B compares the costs of schedule delay. Our estimates are of the same order of magnitude as those of \citet{kobayashi2020J.JSCESer.D3} for urban rail commuters in Tokyo and \citet{peer2016TransportationResearchPartB:Methodological} for rail commuters in the Netherlands, both based on revealed-preference data. For both early and late arrival, our estimates lie between theirs. By contrast, our estimates are substantially smaller than those of \citet{xu2024TransportationResearchPartB:Methodological}, whose modeling framework is broadly similar to ours. This variation across studies suggests that the valuation of schedule delay depends substantially on modeling assumptions, in particular on how desired arrival times are specified.

The finding that schedule delay is more costly than crowding holds in all specifications that allow for heterogeneity in desired CBD arrival times, although the levels of crowding and late-arrival costs depend on how desired arrival times are specified (Section~\ref{subsec:heterogeneity}). In urban rail systems with chronic peak-period crowding, such as those in the Tokyo metropolitan area, commuter-side incentives alone may therefore have limited effectiveness in flattening peak crowding under a given distribution of work start times, and crowding mitigation should address not only commuters' train choices but also the temporal constraints that shape those choices. Off-peak discounts, off-peak point incentives, and the provision of crowding information are designed to alter train choice for a given desired arrival time. By contrast, staggered work start times and flexible working-hour arrangements can make desired arrival times themselves more flexible, which may facilitate shifts to less crowded trains and help smooth peak-period crowding. Combining policies that influence commuters' train choices with policies that relax the underlying time constraints may therefore be important for effective crowding mitigation. Because crowding is determined in equilibrium in our model, the estimated preferences can be used to evaluate such policies while accounting for the resulting adjustment of crowding. We are conducting such counterfactual analyses as part of ongoing work.

\section{Conclusion}
\label{sec:conclusion}

This study estimates commuters' preferences for schedule delay early, schedule delay late, travel time, and crowding in urban rail, accounting for the interaction between train choice and crowding. We use individual-level data from the 2015 PTSMA, focusing on the Seibu Shinjuku Line in the Tokyo metropolitan area. We model the interaction between commuters' train choices and crowding as a potential game, which allows us to estimate preference parameters under an equilibrium with endogenous crowding and evaluate their statistical uncertainty. The survey data allow us to construct desired arrival times mainly from commuters' reported work start times, together with an assumed preferred buffer time, rather than from their observed train choices.

The main finding is that commuters prefer to avoid crowding but place greater weight on avoiding deviations from their desired arrival times. In the baseline estimation, all four cost parameters are positive, and their 95\% confidence intervals exclude zero. One minute of schedule delay early and late corresponds to 0.39 and 0.49 minutes of in-vehicle travel time, respectively, whereas one minute of travel at a crowding rate of 100\% adds a crowding cost equivalent to 0.04 minutes. This ranking holds when we allow for heterogeneity in desired arrival times, although the crowding parameter is sensitive to this specification and reaches zero in some cases. Commuters may therefore rationally choose crowded peak-hour trains to avoid arriving earlier or later than desired, which helps explain why peak-hour crowding persists in equilibrium.

These findings suggest that policies to reduce crowding should account for commuters' scheduling constraints, for example by combining incentives to change trains with staggered work start times or flexible working hours. We will quantify the effects of such policies, accounting for the equilibrium adjustment of crowding, through counterfactual simulations in a future version.

This study has several limitations that point to directions for future research. First, we assume that commuters know other commuters' desired arrival times and the crowding on each train. Incorporating incomplete information would allow an analysis of information-based policies such as real-time crowding information. Second, we infer the train each commuter used by matching reported times with the timetable, so train choices and the resulting crowding levels may contain measurement error, which could bias the estimated crowding costs.\footnote{In the current version, some constructed crowding rates are implausibly high (Figure~\ref{fig:app-crowding}), which likely reflects such matching errors. We plan to refine the train-matching procedure in a future version.} Third, crowding disutility is linear in the crowding rate and does not distinguish seated from standing travel. Nonlinear specifications could capture the additional disutility of severe crowding suggested by the external benchmarks. Fourth, the model abstracts from limited express trains and from transfers between trains on the line, both of which could affect train choices in practice.

The remaining limitations concern statistical inference. Our parametric bootstrap treats the expanded commuter population as given, so its standard errors do not reflect uncertainty from survey sampling or from the construction of the expansion weights and may understate the overall uncertainty. A survey-based bootstrap that resamples individual survey records and recomputes the expansion weights would capture these sources of uncertainty \citep{rao1988JournaloftheAmericanStatisticalAssociation,rao1992SurveyMethodology}. In addition, we treat station--train cells with no observed commuters as zeros, although some of these cells may have positive demand that the survey did not capture. Smoothing or hierarchical models that borrow information from related stations and trains, or a likelihood that explicitly models the survey observation process, could address such cells, although these approaches require additional assumptions about train choices in cells with no observations.

Beyond these limitations, two extensions would broaden the scope of the analysis. First, extending the model from a single line to the rail network as a whole would allow us to model commuters' route choices across lines and to reduce the share of passengers treated as exogenous. Second, incorporating the choices of work start times \citep{henderson1981JournalofUrbanEconomics,takayama2015TransportationResearchPartB:Methodological} and of residential and workplace locations \citep{takayama2017JournalofUrbanEconomics,fosgerau2018JournalofUrbanEconomics}, which we take as given, would allow an analysis of how crowding and scheduling costs shape when people work and where they live and work, and of the long-run effects of policies such as staggered work start times.

\appendix
\counterwithin{figure}{section}
\counterwithin{table}{section}
\section*{Appendix}

\section{Data construction and additional descriptive evidence}
\label{app:data}

This appendix describes the PTSMA survey and its expansion factors (Section~\ref{app:ptsma-expansion}), the construction of train-level data, including train matching, the allocation of expanded counts across trains, and train capacities (Section~\ref{app:train-level-data}), and additional descriptive figures (Section~\ref{app:additional-descriptive}).

\subsection{PTSMA survey and expansion factors}
\label{app:ptsma-expansion}

The 2015 PTSMA railway passenger survey was conducted from November 17 to 19, 2015, with the survey on the Seibu Shinjuku Line conducted on November 17. Survey staff distributed questionnaires to passengers getting off trains at selected railway stations. Respondents could return their answers by mail, at designated collection points, or online. The questionnaire collected information on railway trips throughout the day, including return trips home.

The PTSMA microdata come with two sets of expansion factors based on different population benchmarks. The first uses commuter pass sales by sales office, sex, and age, and applies only to commuting and school trips made with commuter passes, with one factor per respondent. The second uses station-to-station passenger counts from automatic ticket gates and applies to all railway trips regardless of purpose or ticket type, with one factor per trip. We use the first set for the players in our model and for the other commuter-pass users, who enter the model as exogenous passengers. We use the second set for the remaining exogenous passengers, such as those traveling with ordinary tickets or for non-commuting purposes, after excluding respondents already covered by the first set.

\subsection{Construction of train-level data}
\label{app:train-level-data}

We match each PTSMA respondent to a scheduled train using their boarding and alighting stations and times. When the boarding or alighting time on the Seibu Shinjuku Line is not recorded because the trip uses several railway lines, we impute it as described in Section~\ref{subsec:ptsma}. When a respondent does not report a boarding or alighting time, we compute it from the reported departure time and access time, or from the reported arrival time and egress time. For respondent $i$, let $\mathcal{M}_i$ denote the set of trains that stop at both their boarding and alighting stations on the Seibu Shinjuku Line. For $m\in\{\mathrm{a},\mathrm{b}\}$, denoting alighting and boarding, let $\omega_i^m$ be the respondent's alighting or boarding time and $\sigma_r^m$ the scheduled arrival ($m=\mathrm{a}$) or departure ($m=\mathrm{b}$) time of train $r$ at the corresponding station. We define
\begin{equation*}
q_i^m =
\begin{cases}
1 & \text{if $\omega_i^m$ is reported},\\
2 & \text{if $\omega_i^m$ is computed from access or egress times},\\
3 & \text{if $\omega_i^m$ is imputed from fare-calculation distances (Section~\ref{subsec:ptsma})}.
\end{cases}
\end{equation*}
With $d_{ir}^m=|\sigma_r^m-\omega_i^m|$, we consider the set of admissible comparisons
\begin{equation*}
\mathcal{C}_i=\{(r,m): r\in\mathcal{M}_i,\ \omega_i^m \text{ is available},\ d_{ir}^m\le 10\}.
\end{equation*}
Each element $(r,m)$ compares the scheduled time of train $r$ with the respondent's alighting or boarding time, so the same train can appear twice in $\mathcal{C}_i$. The matched train $\hat r_i$ solves
\begin{equation}
(\hat r_i,\hat m_i)=\operatorname*{lexmin}_{(r,m)\in\mathcal{C}_i}\Bigl(q_i^m,\ \mathbf{1}\{d_{ir}^m>0\},\ \mathbf{1}\{m=\mathrm{b}\},\ d_{ir}^m,\ \mathbf{1}\{\sigma_r^m>\omega_i^m\}\Bigr),
\end{equation}
where $\operatorname{lexmin}$ selects the smallest vector in lexicographic order. That is, vectors are first compared by their first components. If the first components are tied, they are compared by their second components, and so on. Thus, we prefer reported times to computed times and computed times to imputed times, exact matches to approximate ones, alighting times to boarding times, smaller time differences, and, among equally close trains, those scheduled at or before the respondent's time. The timetable we use covers only trains that do not require a surcharge, so respondents who rode limited express trains are matched to other trains. The PTSMA does not record transfers between trains on the line in a consistent way: some respondents report each train separately, whereas others report a single trip from their boarding station to their alighting station. We use only the first boarding station and the last alighting station on the line, so respondents who made such transfers are matched to a single train.

Because the expansion factors are not constructed at the train level, applying them directly to the matched trains could produce erratic train-level counts. We therefore use the expansion factors only to compute totals and allocate each total across trains in proportion to the number of respondents matched to each train, as in Section~\ref{subsec:expansion}. For the commuter-pass-based factors, the totals are computed for each boarding station on the Seibu Shinjuku Line. For the automatic-ticket-gate-based factors, which are defined for station-to-station flows, they are computed for each pair of boarding and alighting stations. Each exogenous respondent therefore represents the total of their station or station pair divided by the number of respondents in it. We add this number to every section between the respondent's boarding and alighting stations on the matched train, and summing over respondents by train and section gives the exogenous passenger counts $E_r^k$.

To derive train-level capacities from the period-level capacity data described in Section~\ref{subsec:timetable}, we use a capacity of 140 passengers per car. Trains on the Seibu Shinjuku Line run with either eight or ten cars, which corresponds to 1,120 passengers for an eight-car train and 1,400 passengers for a ten-car train. Kami-Igusa, Iogi, and Shimo-Igusa are served only by local trains and cannot accommodate ten-car trains. All local trains in our analysis run to the CBD and therefore stop at these stations, so they are assigned eight cars. For the other trains, we choose the number of cars so that the total capacity of the trains passing through each section in each time period matches the PTSMA data. This determines the number of cars uniquely except for the semi-express train arriving at the CBD at 9:12 and the express train arriving at 9:16, one of which must have eight cars. Because the available information does not indicate which, we assign ten cars to the express train, which makes fewer stops, and eight cars to the semi-express train. As a result, all trains other than local trains have ten cars except this semi-express train.

\subsection{Additional descriptive evidence}
\label{app:additional-descriptive}

\begin{sidewaysfigure}[p]
    \centering
    \includegraphics[width=\textheight]{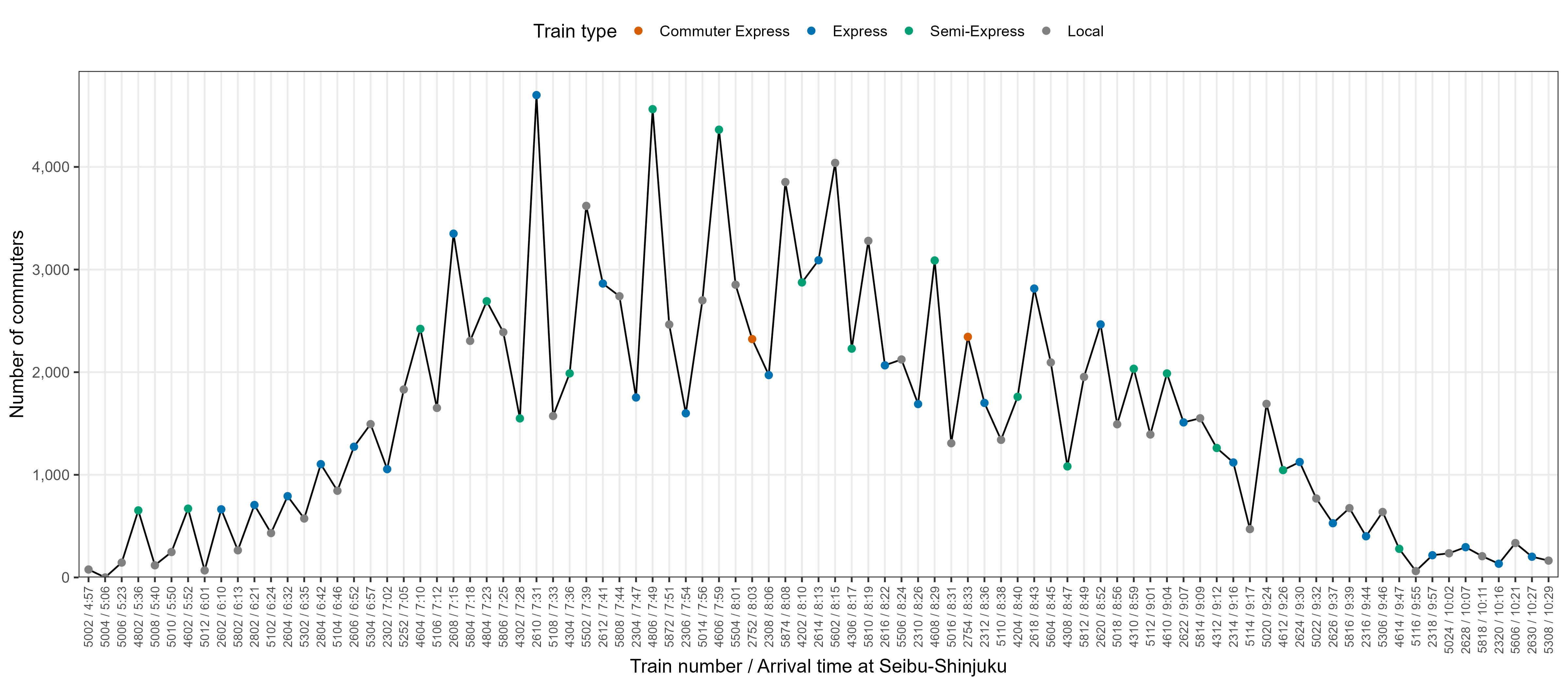}

    \caption{Distribution of players across trains}
    \label{fig:app-players}

    \vspace{0.4em}
    \begin{minipage}{0.95\textheight}
        \footnotesize
        \textit{Notes:} Players are commuters whose train choices
        are modeled in the analysis. The figure shows their numbers
        on each train along the Shimo-Ochiai--Takadanobaba section,
        approaching the CBD. Train arrival times are measured at
        Seibu-Shinjuku Station, the CBD terminal.
    \end{minipage}
\end{sidewaysfigure}

\begin{sidewaysfigure}[p]
    \centering
    \includegraphics[width=\textheight]{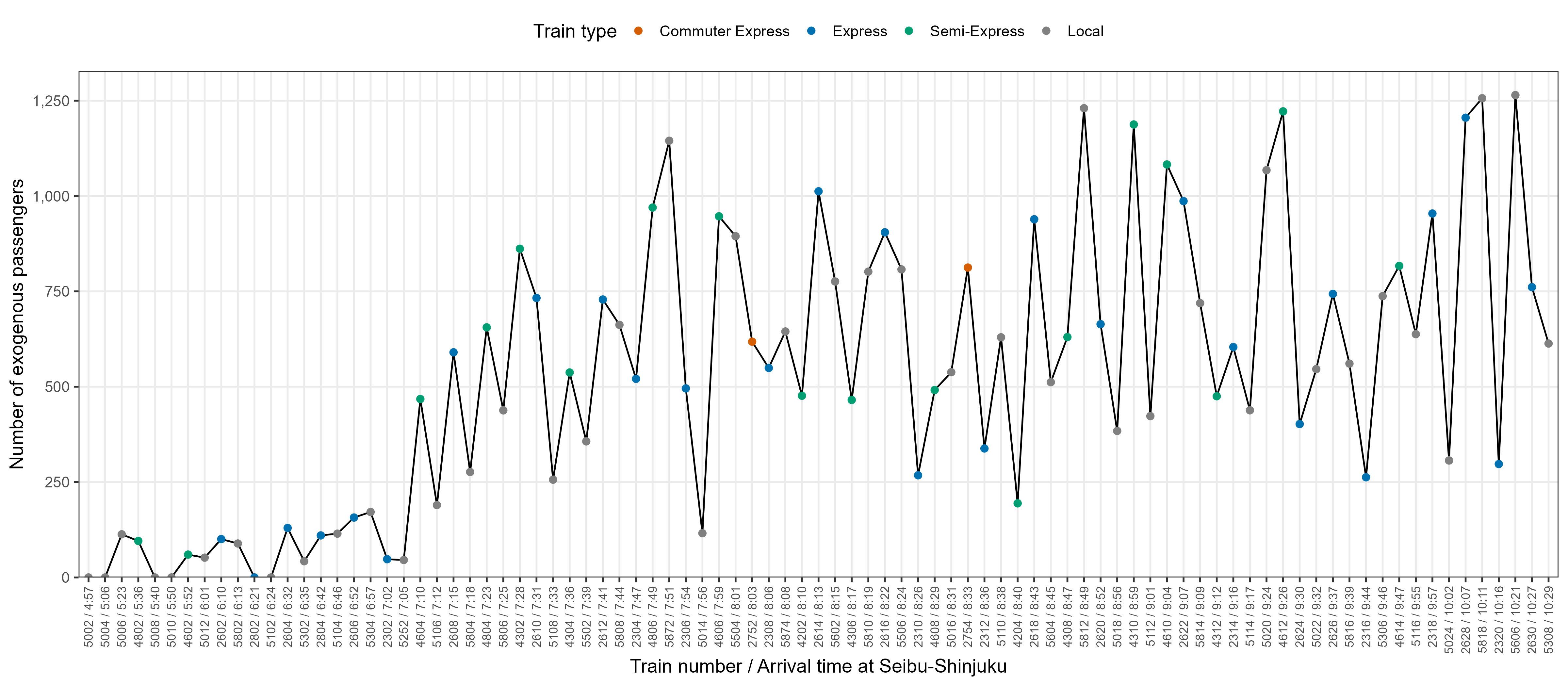}

    \caption{Distribution of exogenous passengers across trains}
    \label{fig:app-exogenous}

    \vspace{0.4em}
    \begin{minipage}{0.95\textheight}
        \footnotesize
        \textit{Notes:} Exogenous passengers are passengers whose
        train choices are not modeled in the analysis. The figure
        shows their estimated numbers on each train along the
        Shimo-Ochiai--Takadanobaba section, approaching the CBD.
        Train arrival times are measured at Seibu-Shinjuku Station,
        the CBD terminal.
    \end{minipage}
\end{sidewaysfigure}

\begin{sidewaysfigure}[p]
    \centering
    \includegraphics[width=\textheight]{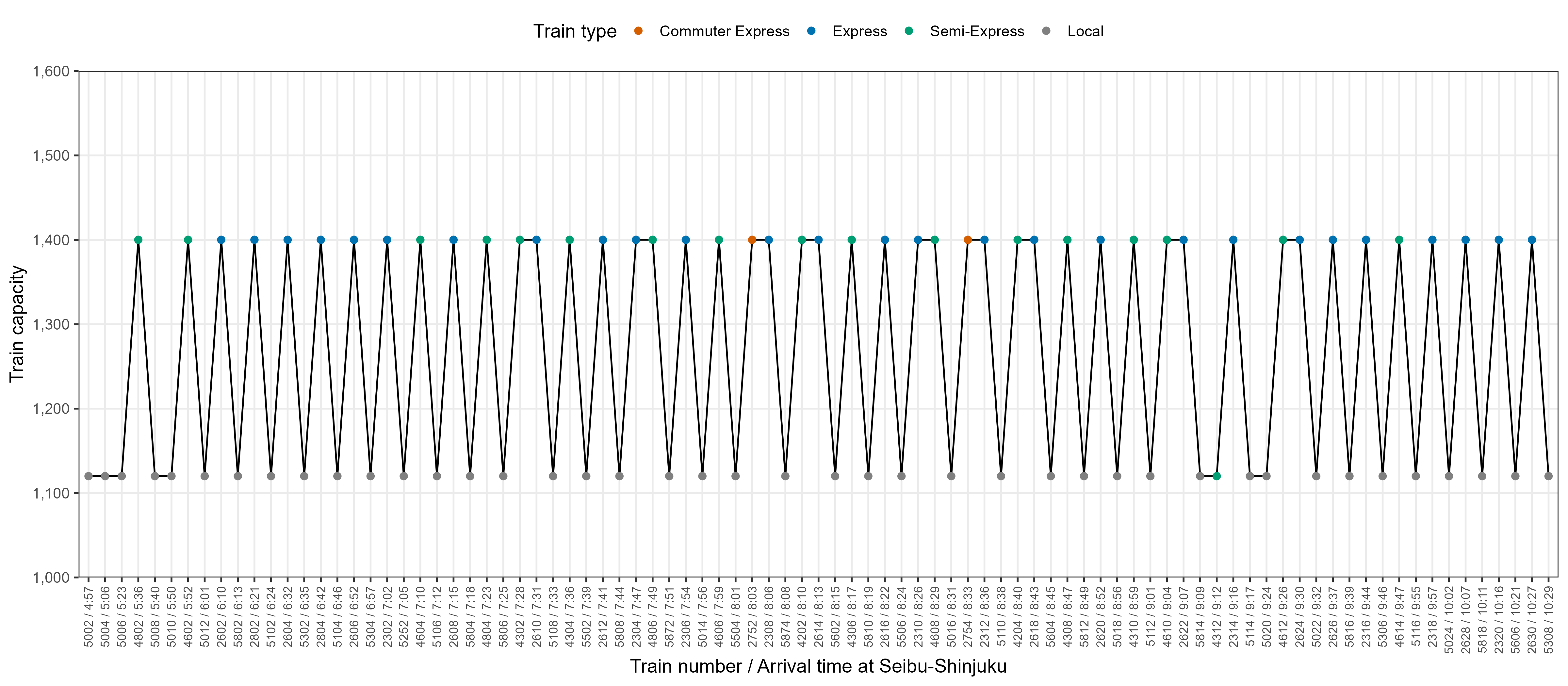}

    \caption{Passenger capacities by train}
    \label{fig:app-capacity}

    \vspace{0.4em}
    \begin{minipage}{0.95\textheight}
        \footnotesize
        \textit{Notes:} Capacity is the rated passenger capacity
        of each train, which counts both seats and
        standing room. Train arrival times are measured at
        Seibu-Shinjuku Station, the CBD terminal. Colors indicate
        train types.
    \end{minipage}
\end{sidewaysfigure}

\begin{sidewaysfigure}[p]
    \centering
    \includegraphics[width=\textheight]{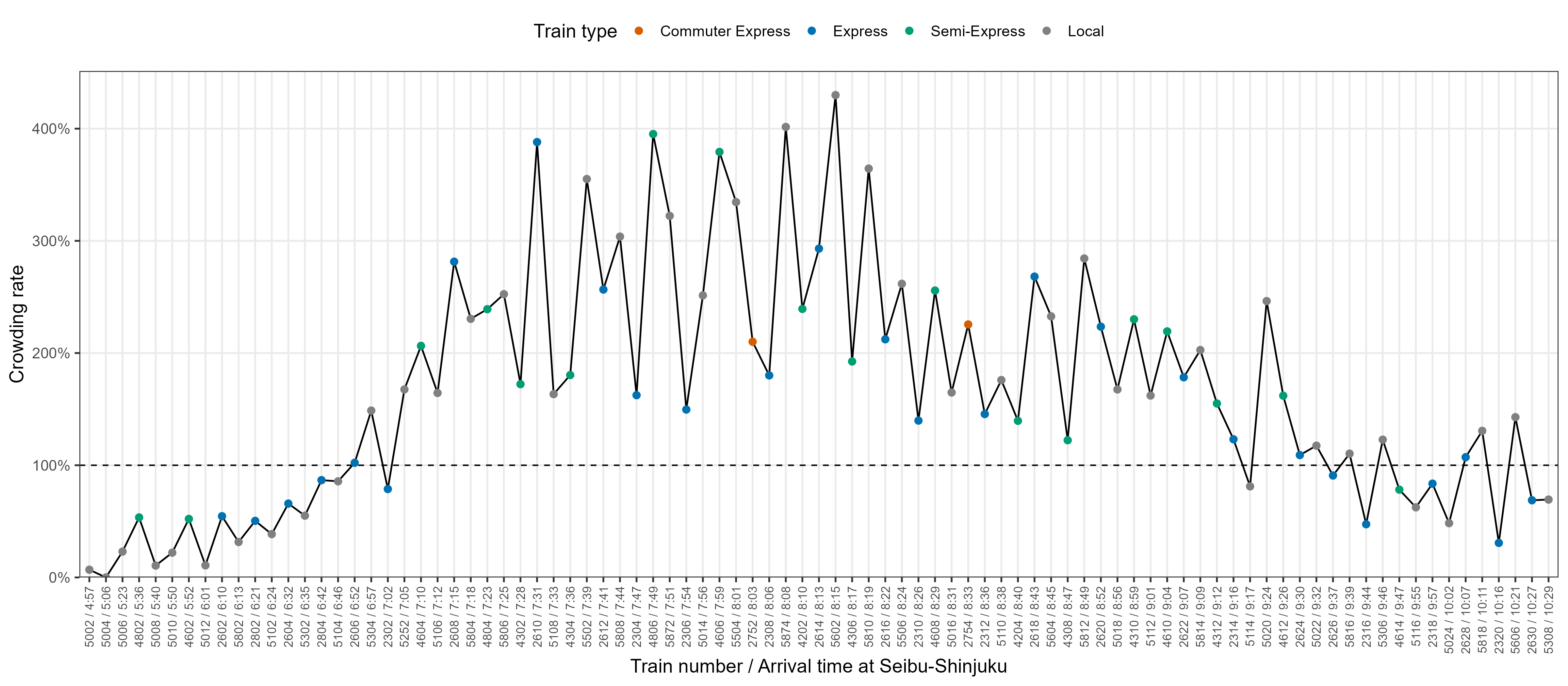}

    \caption{Crowding rates by train}
    \label{fig:app-crowding}

    \vspace{0.4em}
    \begin{minipage}{0.95\textheight}
        \footnotesize
        \textit{Notes:} Crowding rates are calculated as the sum of
        players and exogenous passengers divided by train capacity.
        Exogenous passenger counts are measured along the
        Shimo-Ochiai--Takadanobaba section, approaching the CBD.
        Train arrival times are measured at Seibu-Shinjuku Station,
        the CBD terminal. The dashed line indicates a crowding
        rate of 100\%.
    \end{minipage}
\end{sidewaysfigure}

Figures~\ref{fig:app-players}--\ref{fig:app-crowding} show train-level quantities on the Shimo-Ochiai--Takadanobaba section, the last section before the CBD. The number of players rises steeply after 7:00, peaks between about 7:30 and 8:20, and declines after 9:00, with large differences even between consecutive trains (Figure~\ref{fig:app-players}). Exogenous passengers are fewer than players during the peak but remain at a similar level after 9:00, so they account for a larger share of passengers later in the morning (Figure~\ref{fig:app-exogenous}). Train capacities alternate between 1,120 passengers for eight-car local trains and 1,400 passengers for ten-car trains, as described in Section~\ref{app:train-level-data} (Figure~\ref{fig:app-capacity}). As a result, crowding rates largely follow the number of players and exceed 100\% for most trains arriving between about 7:00 and 9:30, with some exceeding 300\% (Figure~\ref{fig:app-crowding}). Such high values are likely to reflect errors in train matching, as noted in Section~\ref{sec:conclusion}.

\begin{figure}[p]
    \centering
    \includegraphics[
        width=0.95\textwidth,
        height=0.80\textheight,
        keepaspectratio
    ]{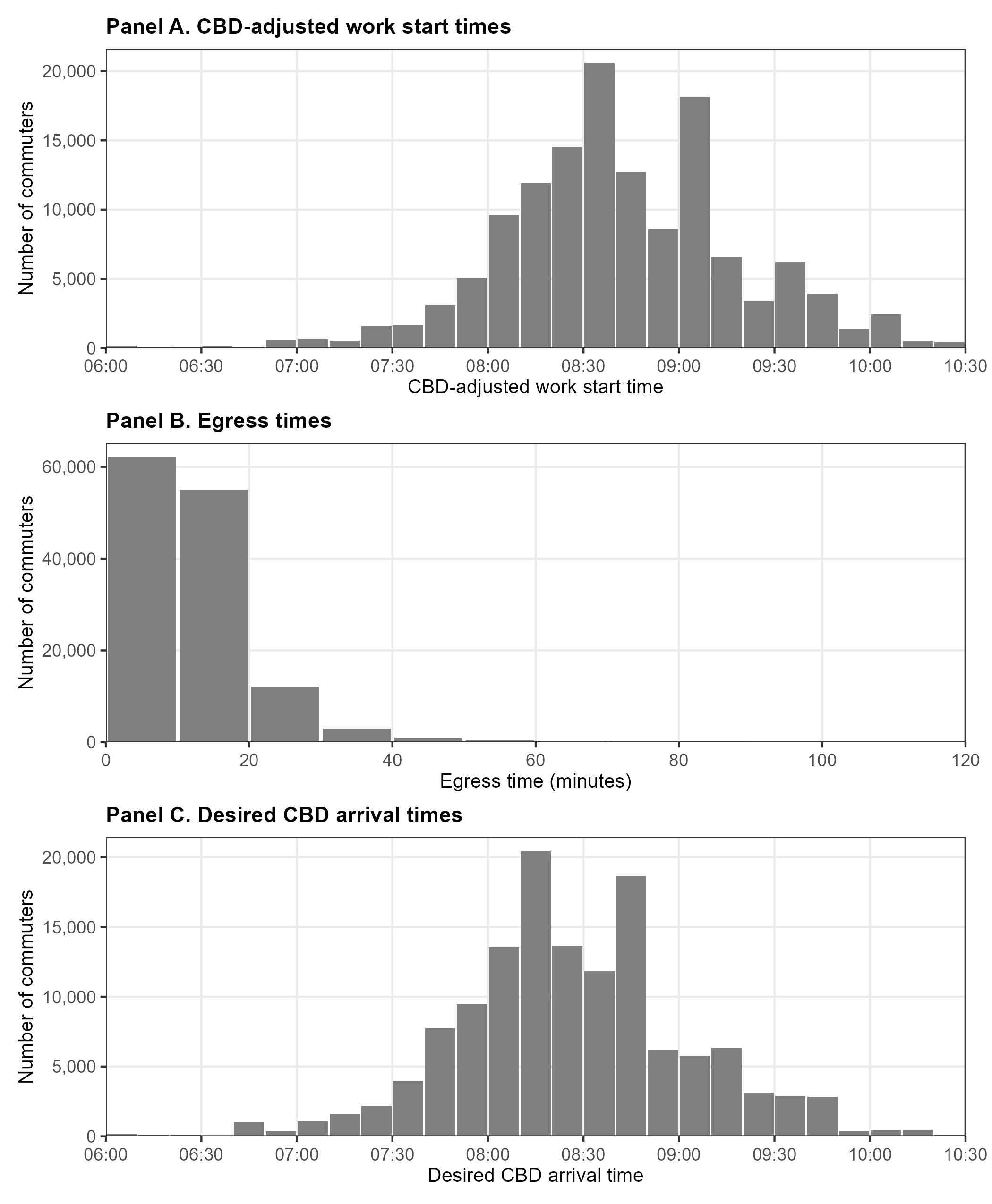}

    \caption{Distributions of work start times, egress times, and desired CBD arrival times}
    \label{fig:app-desired-baseline}

    \vspace{0.4em}
    \begin{minipage}{0.95\textwidth}
        \footnotesize
        \textit{Notes:} Panel A shows CBD-adjusted work start times,
        calculated by subtracting the railway travel time after the
        CBD, if any, from the reported work start time. Panel B shows egress times from the final railway
        station to the workplace. Panel C shows desired CBD arrival
        times, calculated by subtracting egress time and a preferred
        buffer time of 15 minutes from CBD-adjusted work start time.
        All distributions use 10-minute intervals.
    \end{minipage}
\end{figure}

\begin{figure}[p]
    \centering
    \includegraphics[
        width=0.95\textwidth,
        height=0.80\textheight,
        keepaspectratio
    ]{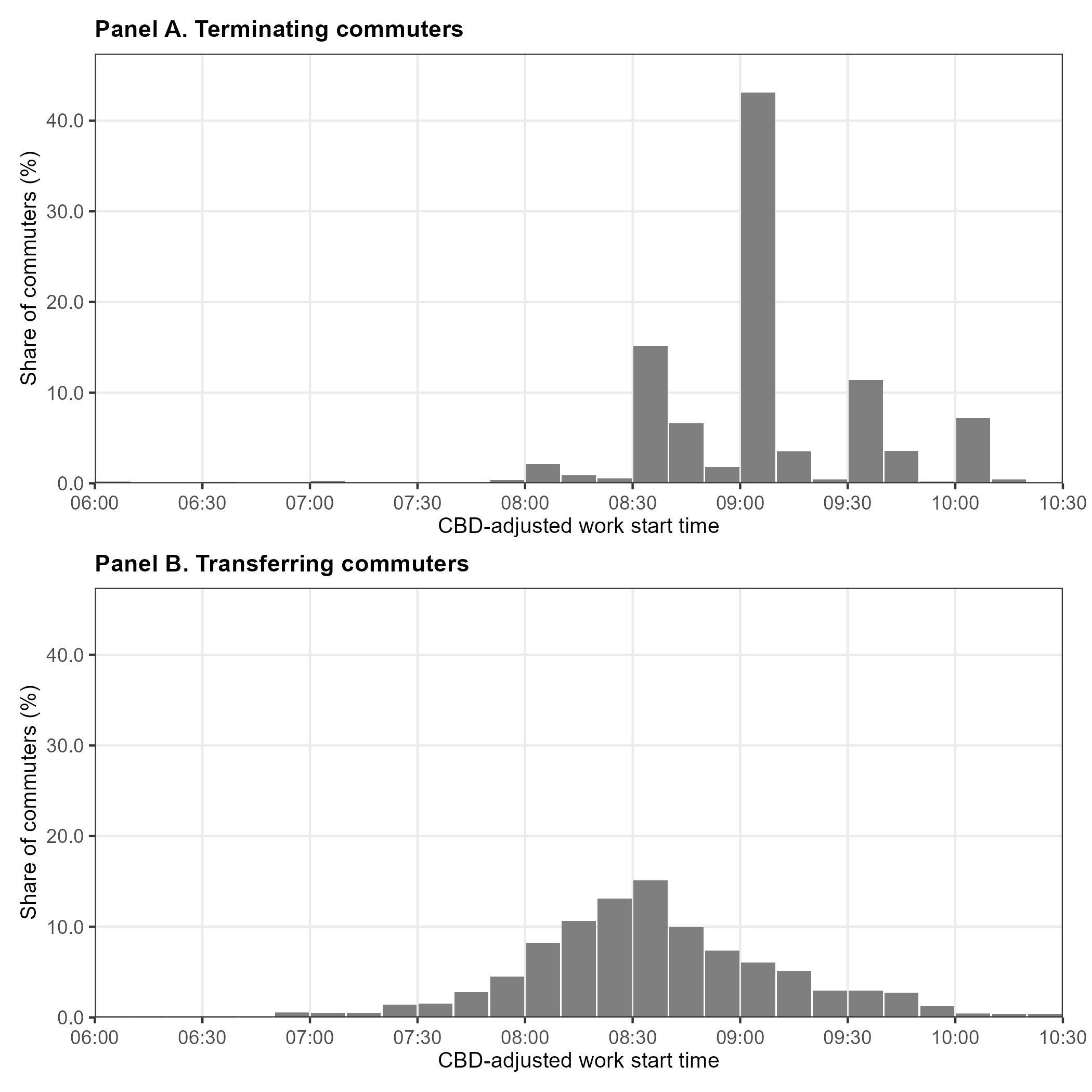}

    \caption{Distributions of CBD-adjusted work start times by transfer status}
    \label{fig:app-work-start-transfer}

    \vspace{0.4em}
    \begin{minipage}{0.95\textwidth}
        \footnotesize
        \textit{Notes:} Panel A shows terminating commuters, and
        Panel B shows transferring commuters. CBD-adjusted work
        start time is calculated by subtracting the railway travel
        time after the CBD, if any, from the reported work start
        time. The distributions use 10-minute
        intervals and show percentages within each commuter group.
    \end{minipage}
\end{figure}

\begin{figure}[p]
    \centering
    \includegraphics[
        width=0.95\textwidth,
        height=0.80\textheight,
        keepaspectratio
    ]{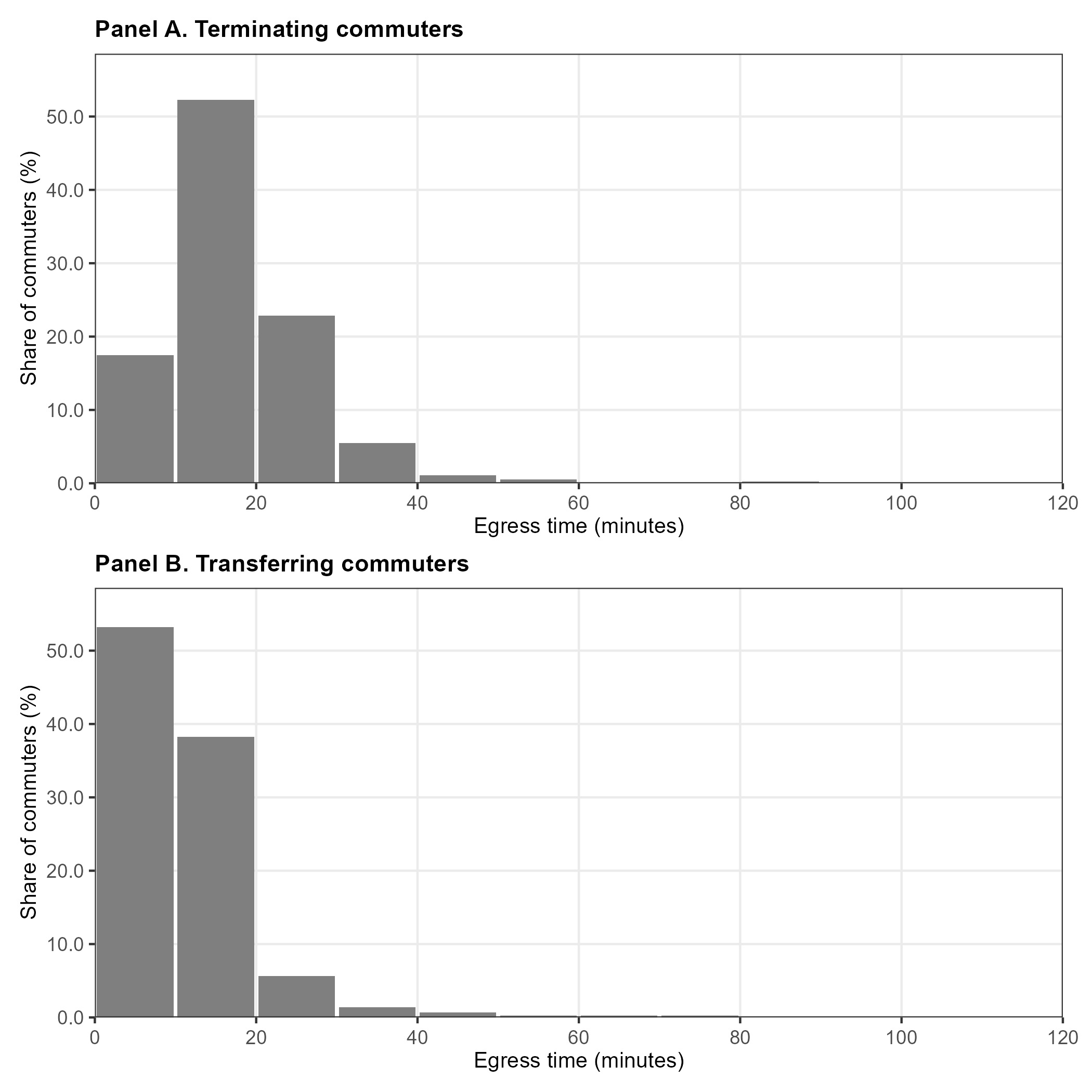}

    \caption{Distributions of egress times by transfer status}
    \label{fig:app-egress-transfer}

    \vspace{0.4em}
    \begin{minipage}{0.95\textwidth}
        \footnotesize
        \textit{Notes:} Panel A shows terminating commuters, and
        Panel B shows transferring commuters. Egress time is measured
        as the difference between arrival time at the workplace and
        departure time from the final railway station. The
        distributions use 10-minute intervals and show percentages
        within each commuter group.
    \end{minipage}
\end{figure}

\begin{figure}[p]
    \centering
    \includegraphics[
        width=0.95\textwidth,
        height=0.80\textheight,
        keepaspectratio
    ]{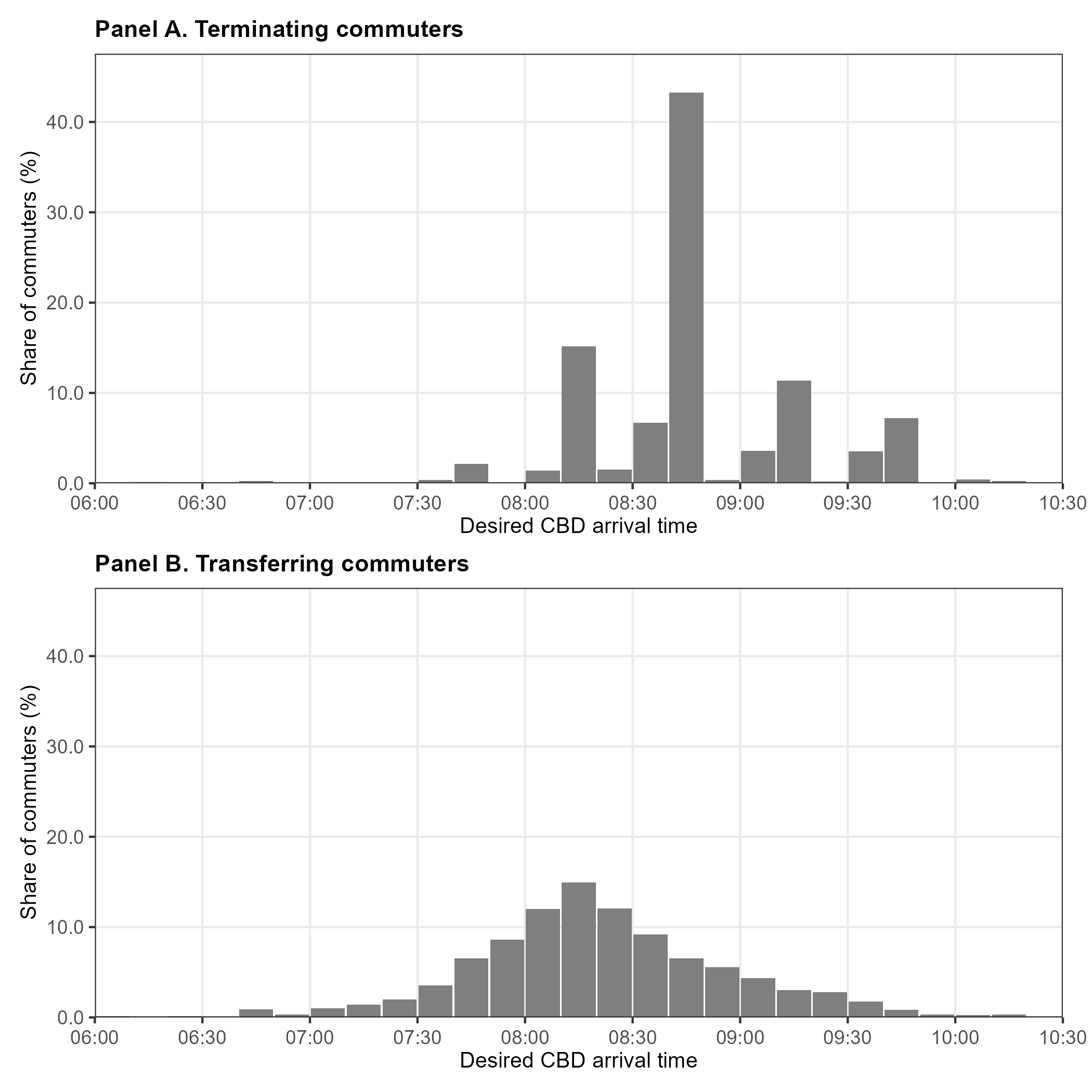}

    \caption{Distributions of desired CBD arrival times by transfer status}
    \label{fig:app-desired-transfer}

    \vspace{0.4em}
    \begin{minipage}{0.95\textwidth}
        \footnotesize
        \textit{Notes:} Panel A shows terminating commuters, and
        Panel B shows transferring commuters. Desired CBD arrival
        time is calculated by subtracting egress time and a preferred
        buffer time of 15 minutes from CBD-adjusted work start time.
        The distributions use 10-minute intervals and show
        percentages within each commuter group.
    \end{minipage}
\end{figure}

Figure~\ref{fig:app-desired-baseline} shows that CBD-adjusted work start times are concentrated between 8:00 and 9:15, with peaks at 8:30 and 9:00, and that most egress times are below 20 minutes. Desired CBD arrival times are accordingly concentrated between about 7:50 and 9:00, around the baseline value of $t^*=8{:}10$. Figures~\ref{fig:app-work-start-transfer}--\ref{fig:app-desired-transfer} show that terminating commuters tend to start work at round times such as 9:00 and have longer egress times, whereas transferring commuters have more dispersed work start times and shorter egress times. The desired CBD arrival times of terminating commuters are therefore later, which motivates the separate values of $t^*$ in Section~\ref{subsubsec:heterogeneity-transfer}.

\section{Numerical optimization}
\label{app:numerical-optimization}

Evaluation of the likelihood in
\eqref{eq:log-likelihood_function} requires a nested optimization, because
the likelihood involves the maximum of the potential over $\mathcal X$.  In
the inner problem, for each candidate parameter vector $\theta$, we maximize
the potential:
\begin{equation}
    \bm x^\ast(\theta)
    = \arg\max_{\bm x\in\mathcal X}\Phi^\ast(\bm x;\theta),
\end{equation}
whose solution is unique because $\Phi^\ast$ is strictly concave in $\bm x$.
In the outer problem, we minimize the negative log-likelihood
\begin{equation}
    -\ell(\theta\mid\bm x^{\mathrm{obs}})
    =-\Phi^\ast(\bm x^{\mathrm{obs}};\theta)
      +\Phi^\ast(\bm x^\ast(\theta);\theta)
\end{equation}
over $\theta$ subject to the nonnegativity restrictions on the cost
parameters.  Appendices~\ref{app:inner-optimization} and
\ref{app:outer-optimization} describe the inner and outer problems,
respectively.

\subsection{Inner potential maximization}
\label{app:inner-optimization}

The inner problem maximizes $\Phi^\ast$ over a product of
population-specific simplexes.  We solve it in two stages: mirror ascent,
which operates directly on the simplexes, first brings the allocation close
to the maximizer, and L-BFGS-B in reduced softmax coordinates then refines
it.  Throughout, let $\psi_r^\ell=\partial\Phi^\ast/\partial x_r^\ell$ and
$p_r^\ell=x_r^\ell/s^\ell$.  The inner problem maximizes
$\Phi^\ast$ subject to the adding-up constraints
$\sum_{r\in\mathcal R^\ell}x_r^\ell=s^\ell$ and nonnegativity.  Because the
derivative of the entropy term diverges as $x_r^\ell\to0$, the maximizer is
interior, and the KKT conditions reduce to
\begin{equation}
    \psi_r^\ell=\mu^\ell,
    \qquad \ell\in[L],\ r\in\mathcal R^\ell,
\end{equation}
where $\mu^\ell$ is the Lagrange multiplier on the adding-up constraint of
population $\ell$.  That is, all feasible strategies in a population have
the same marginal potential.  Since $\sum_{q\in\mathcal R^\ell}p_q^\ell=1$,
the multiplier $\mu^\ell$ equals the share-weighted average
$\sum_{q\in\mathcal R^\ell}p_q^\ell\psi_q^\ell$ at the maximizer.  We
therefore measure the violation of the KKT conditions at any allocation by
\begin{equation}
    R_{\mathrm{KKT}}
    =\max_{\ell}\max_{r\in\mathcal R^\ell}
      \left|
      \psi_r^\ell-
      \sum_{q\in\mathcal R^\ell}p_q^\ell \psi_q^\ell
      \right|,
    \label{eq:inner-kkt-residual}
\end{equation}
which is zero at the maximizer and is used to assess convergence.  Since
$\Phi^\ast$ is strictly concave in $\bm x$ (Section~\ref{subsec:likelihood}),
the KKT conditions are also sufficient, and the maximizer is unique.

In the first stage, we use mirror ascent with negative entropy as the mirror
map \citep[Section~4.3]{bubeck2015}.  Given a step size $a_m$, the
within-population shares are updated according to
\begin{equation}
    p_{r,m+1}^\ell
    =
    \frac{p_{r,m}^\ell\exp(a_m \psi_{r,m}^\ell)}
    {\sum_{q\in\mathcal R^\ell}
      p_{q,m}^\ell\exp(a_m \psi_{q,m}^\ell)},
\end{equation}
where $\psi_{r,m}^\ell$ is $\psi_r^\ell$ evaluated at the $m$-th iterate.
This update keeps every share positive and every population total $s^\ell$
fixed, so each iterate is feasible.  The step size is chosen by
backtracking so that the potential increases at each iteration.  The first
stage stops when $R_{\mathrm{KKT}}$ falls below $10^{-6}$ or after 1,000
iterations.

In the second stage, we refine the resulting allocation using the
limited-memory BFGS method with bound constraints (L-BFGS-B), as implemented
in SciPy's \texttt{minimize}.  To impose the adding-up and positivity
restrictions, we use reduced softmax coordinates.  Because adding the same
constant to every logit in a population leaves the allocation unchanged, we
fix the logit of one feasible reference strategy $r_0^\ell$ in each
population at zero and use the remaining $|\mathcal R^\ell|-1$ logits as
free coordinates:
\begin{align}
    x_r^\ell
    &=s^\ell
      \frac{\exp(z_r^\ell)}
      {1+\sum_{q\in\mathcal R^\ell\setminus\{r_0^\ell\}}
       \exp(z_q^\ell)},
       &&r\ne r_0^\ell,\\
    x_{r_0^\ell}^\ell
    &=s^\ell
      \frac{1}
      {1+\sum_{q\in\mathcal R^\ell\setminus\{r_0^\ell\}}
       \exp(z_q^\ell)}.
\end{align}
Every interior allocation corresponds to a unique vector of reduced logits.
The analytic gradient of the negative
potential with respect to a nonreference logit is
\begin{equation}
    \frac{\partial[-\Phi^\ast]}{\partial z_r^\ell}
    =-x_r^\ell
      \left(\psi_r^\ell-\sum_{q\in\mathcal R^\ell}p_q^\ell \psi_q^\ell\right).
    \label{eq:reduced-logit-gradient}
\end{equation}
Starting from the logits corresponding to the first-stage allocation, we
minimize the negative potential over the reduced logits and supply the
analytic gradient in \eqref{eq:reduced-logit-gradient}.  Because the reduced
logits are unconstrained, no bounds are imposed in the inner problem.  The
tolerances are \texttt{ftol}$=10^{-12}$ and \texttt{gtol}$=10^{-9}$, with at
most 1,000 iterations.

Mirror ascent thus supplies a stable
starting point, while L-BFGS-B provides the final local refinement.  For
each inner solution, we record both the optimizer's success flag and
$R_{\mathrm{KKT}}$.  Agreement between the two-stage solver and L-BFGS-B
applied without the first stage provides a further check that the reported
result is not an artifact of the simplex parameterization.

\subsection{Outer parameter optimization and diagnostics}
\label{app:outer-optimization}

The outer problem is solved by L-BFGS-B with lower bounds of zero.  Its
analytic gradient follows from the envelope theorem:
\begin{equation}
    \nabla_\theta[-\ell(\theta\mid\bm x^{\mathrm{obs}})]
    =-\partial_\theta\Phi^\ast(\bm x^{\mathrm{obs}};\theta)
      +\partial_\theta\Phi^\ast(\bm x^\ast(\theta);\theta).
    \label{eq:outer-envelope-gradient}
\end{equation}
Consequently, derivatives of $\bm x^\ast(\theta)$ need not be computed.
The outer tolerances are \texttt{ftol}$=10^{-12}$ and
\texttt{gtol}$=10^{-5}$, with at most 500 iterations.  The gradient tolerance
is looser than in the inner problem because the envelope gradient inherits
numerical error from the inner solution.  For the same reason, although the
log-likelihood is concave in $\theta$, so that any local maximum is global,
the final estimates should also be checked using multiple starting values
and, where useful, a coarse likelihood grid.

\section{Detailed estimation and bootstrap results}
\label{app:detailed-estimation}

Tables~\ref{tab:bootstrap-results}--\ref{tab:bootstrap-60min} report detailed bootstrap results for the baseline specification and the two alternative desired CBD arrival times (Table~\ref{tab:bootstrap-results}), the specification with desired CBD arrival times differing by transfer status (Table~\ref{tab:bootstrap-transfer}), and the 30-minute and 60-minute groupings of CBD-adjusted work start times with and without the late-arrival dummy (Tables~\ref{tab:bootstrap-30min} and~\ref{tab:bootstrap-60min}). In all specifications, the bootstrap means are close to the corresponding point estimates, and the estimated biases are small.

\begin{table}[htbp]
\centering
\caption{Bootstrap results: baseline}
\label{tab:bootstrap-results}
\small
\resizebox{\textwidth}{!}{%
\begin{tabular}{llrrrrr}
\toprule
Desired CBD arrival time $t^\ast$ & Parameter & Estimate & Bootstrap mean & Bias & Bootstrap SE & 95\% interval \\
\midrule
8:00 & $\beta_E/\eta$ & 0.020946 & 0.020923 & -0.000023 & 0.000148 & [0.020643, 0.021206] \\
& $\beta_L/\eta$ & 0.019391 & 0.019371 & -0.000019 & 0.000122 & [0.019127, 0.019608] \\
& $\lambda/\eta$ & 0.001942 & 0.001892 & -0.000050 & 0.000168 & [0.001564, 0.002224] \\
& $\delta/\eta$ & 0.044956 & 0.045054 & 0.000098 & 0.000681 & [0.043695, 0.046360] \\
\addlinespace
8:10 (Baseline) & $\beta_E/\eta$ & 0.018039 & 0.018021 & -0.000018 & 0.000128 & [0.017787, 0.018266] \\
& $\beta_L/\eta$ & 0.022466 & 0.022446 & -0.000019 & 0.000136 & [0.022172, 0.022707] \\
& $\lambda/\eta$ & 0.001848 & 0.001803 & -0.000045 & 0.000166 & [0.001480, 0.002169] \\
& $\delta/\eta$ & 0.046220 & 0.046320 & 0.000100 & 0.000647 & [0.045028, 0.047521] \\
\addlinespace
8:20 & $\beta_E/\eta$ & 0.014785 & 0.014767 & -0.000018 & 0.000108 & [0.014567, 0.014979] \\
& $\beta_L/\eta$ & 0.024941 & 0.024917 & -0.000025 & 0.000154 & [0.024627, 0.025215] \\
& $\lambda/\eta$ & 0.000988 & 0.000940 & -0.000048 & 0.000160 & [0.000645, 0.001277] \\
& $\delta/\eta$ & 0.046443 & 0.046571 & 0.000127 & 0.000668 & [0.045262, 0.047842] \\
\bottomrule
\end{tabular}
}

\vspace{0.5em}
\parbox{0.95\textwidth}{\footnotesize Notes: For the bootstrap, commuter counts in each station--train cell are rounded to the nearest integer, so the point estimates may differ slightly from those in the main tables. The bootstrap SE is the sample standard deviation of the 1,000 retained estimates. All 1,000 bootstrap re-estimations terminated successfully. The 95\% intervals are the 2.5th and 97.5th percentiles of those estimates. These quantities condition on the expanded population and do not include survey-sampling uncertainty. The 8:10 specification is the baseline.}
\end{table}

\begin{table}[htbp]
\centering
\caption{Bootstrap results: heterogeneity by transfer status}
\label{tab:bootstrap-transfer}
\small
\resizebox{\textwidth}{!}{%
\begin{tabular}{llrrrrr}
\toprule
Desired CBD arrival time $t^\ast$ & Parameter & Estimate & Bootstrap mean & Bias & Bootstrap SE & 95\% interval \\
\midrule
8:30 / 8:10 & $\beta_E/\eta$ & 0.017300 & 0.017283 & -0.000017 & 0.000115 & [0.017067, 0.017511] \\
& $\beta_L/\eta$ & 0.024215 & 0.024198 & -0.000018 & 0.000141 & [0.023921, 0.024485] \\
& $\lambda/\eta$ & 0.001535 & 0.001500 & -0.000036 & 0.000157 & [0.001202, 0.001813] \\
& $\delta/\eta$ & 0.046578 & 0.046687 & 0.000109 & 0.000709 & [0.045299, 0.048097] \\
\bottomrule
\end{tabular}
}

\vspace{0.5em}
\parbox{0.95\textwidth}{\footnotesize Notes: The desired CBD arrival time is 8:30 for commuters completing their trips on the study line and 8:10 for commuters transferring to other lines. For the bootstrap, commuter counts in each station--train cell are rounded to the nearest integer, so the point estimates may differ slightly from those in the main tables. The bootstrap SE is the sample standard deviation of the 1,000 retained estimates. All 1,000 bootstrap re-estimations terminated successfully. The 95\% intervals are the 2.5th and 97.5th percentiles of those estimates. These quantities condition on the expanded population and do not include survey-sampling uncertainty.}
\end{table}

\begin{table}[p]
\centering
\caption{Bootstrap results: 30-minute work-start-time groups}
\label{tab:bootstrap-30min}
\scriptsize
\setlength{\tabcolsep}{4pt}
\renewcommand{\arraystretch}{0.9}

\begin{tabular}{llrrrrr}
\toprule
Specification & Parameter & Estimate & Bootstrap mean & Bias & Bootstrap SE & 95\% interval \\
\midrule
\multicolumn{7}{l}{\textit{Panel A: No late dummy, midpoints on the hour and half hour}} \\
\addlinespace
& $\beta_E/\eta$ & 0.023609 & 0.023621 & 0.000012 & 0.000094 & [0.023440, 0.023818] \\
& $\beta_L/\eta$ & 0.076866 & 0.076911 & 0.000045 & 0.000365 & [0.076198, 0.077608] \\
& $\lambda/\eta$ & 0.000000 & 0.000048 & 0.000048 & 0.000081 & [0.000000, 0.000280] \\
& $\delta/\eta$ & 0.048840 & 0.048734 & -0.000107 & 0.000615 & [0.047478, 0.049902] \\
\addlinespace
\multicolumn{7}{l}{\textit{Panel B: No late dummy, midpoints at a quarter past and a quarter to the hour}} \\
\addlinespace
& $\beta_E/\eta$ & 0.025532 & 0.025535 & 0.000002 & 0.000096 & [0.025350, 0.025727] \\
& $\beta_L/\eta$ & 0.077453 & 0.077434 & -0.000018 & 0.000358 & [0.076694, 0.078127] \\
& $\lambda/\eta$ & 0.001364 & 0.001330 & -0.000034 & 0.000153 & [0.001037, 0.001619] \\
& $\delta/\eta$ & 0.052441 & 0.052553 & 0.000112 & 0.000675 & [0.051267, 0.053825] \\
\addlinespace
\multicolumn{7}{l}{\textit{Panel C: With late dummy, midpoints on the hour and half hour}} \\
\addlinespace
& $\beta_E/\eta$ & 0.025569 & 0.025569 & 0.000000 & 0.000105 & [0.025361, 0.025787] \\
& $\beta_L/\eta$ & 0.061827 & 0.061846 & 0.000019 & 0.000399 & [0.061076, 0.062634] \\
& $\mathrm{LateDummy}/\eta$ & 0.540901 & 0.540991 & 0.000090 & 0.009606 & [0.521605, 0.560414] \\
& $\lambda/\eta$ & 0.000277 & 0.000259 & -0.000018 & 0.000139 & [0.000000, 0.000528] \\
& $\delta/\eta$ & 0.049274 & 0.049319 & 0.000044 & 0.000651 & [0.048072, 0.050603] \\
\addlinespace
\multicolumn{7}{l}{\textit{Panel D: With late dummy, midpoints at a quarter past and a quarter to the hour}} \\
\addlinespace
& $\beta_E/\eta$ & 0.027820 & 0.027818 & -0.000001 & 0.000106 & [0.027619, 0.028021] \\
& $\beta_L/\eta$ & 0.062245 & 0.062243 & -0.000002 & 0.000412 & [0.061451, 0.063098] \\
& $\mathrm{LateDummy}/\eta$ & 0.564281 & 0.563681 & -0.000600 & 0.010158 & [0.543672, 0.583171] \\
& $\lambda/\eta$ & 0.002344 & 0.002304 & -0.000040 & 0.000155 & [0.002000, 0.002607] \\
& $\delta/\eta$ & 0.051320 & 0.051420 & 0.000099 & 0.000696 & [0.050079, 0.052738] \\
\bottomrule
\end{tabular}

\vspace{0.5em}
\parbox{0.95\textwidth}{\footnotesize
Notes: The desired CBD arrival times are grouped into 30-minute intervals. Specifications with a late dummy include an additional fixed penalty for arriving after the desired CBD arrival time. For the bootstrap, commuter counts in each station--train cell are rounded to the nearest integer, so the point estimates may differ slightly from those in the main tables. The bootstrap SE is the sample standard deviation of the 1,000 retained estimates. All 1,000 bootstrap re-estimations terminated successfully in each specification. The 95\% intervals are the 2.5th and 97.5th percentiles of those estimates. These quantities condition on the expanded population and do not include survey-sampling uncertainty.
}
\end{table}

\begin{table}[p]
\centering
\caption{Bootstrap results: 60-minute work-start-time groups}
\label{tab:bootstrap-60min}
\scriptsize
\setlength{\tabcolsep}{4pt}
\renewcommand{\arraystretch}{0.9}

\begin{tabular}{llrrrrr}
\toprule
Specification & Parameter & Estimate & Bootstrap mean & Bias & Bootstrap SE & 95\% interval \\
\midrule
\multicolumn{7}{l}{\textit{Panel A: No late dummy, midpoints on the hour}} \\
\addlinespace
& $\beta_E/\eta$ & 0.021844 & 0.021849 & 0.000005 & 0.000089 & [0.021673, 0.022032] \\
& $\beta_L/\eta$ & 0.056427 & 0.056450 & 0.000023 & 0.000257 & [0.055974, 0.056960] \\
& $\lambda/\eta$ & 0.000000 & 0.000041 & 0.000041 & 0.000070 & [0.000000, 0.000230] \\
& $\delta/\eta$ & 0.044251 & 0.044127 & -0.000124 & 0.000548 & [0.043064, 0.045198] \\
\addlinespace
\multicolumn{7}{l}{\textit{Panel B: No late dummy, midpoints on the half hour}} \\
\addlinespace
& $\beta_E/\eta$ & 0.021874 & 0.021886 & 0.000012 & 0.000089 & [0.021710, 0.022061] \\
& $\beta_L/\eta$ & 0.058461 & 0.058467 & 0.000006 & 0.000254 & [0.057939, 0.058970] \\
& $\lambda/\eta$ & 0.000000 & 0.000048 & 0.000048 & 0.000073 & [0.000000, 0.000243] \\
& $\delta/\eta$ & 0.046605 & 0.046494 & -0.000111 & 0.000563 & [0.045339, 0.047578] \\
\addlinespace
\multicolumn{7}{l}{\textit{Panel C: With late dummy, midpoints on the hour}} \\
\addlinespace
& $\beta_E/\eta$ & 0.023020 & 0.023024 & 0.000004 & 0.000099 & [0.022812, 0.023229] \\
& $\beta_L/\eta$ & 0.051198 & 0.051224 & 0.000026 & 0.000296 & [0.050665, 0.051810] \\
& $\mathrm{LateDummy}/\eta$ & 0.255631 & 0.255498 & -0.000133 & 0.008791 & [0.238915, 0.272951] \\
& $\lambda/\eta$ & 0.000000 & 0.000038 & 0.000038 & 0.000066 & [0.000000, 0.000214] \\
& $\delta/\eta$ & 0.044623 & 0.044518 & -0.000105 & 0.000555 & [0.043415, 0.045585] \\
\addlinespace
\multicolumn{7}{l}{\textit{Panel D: With late dummy, midpoints on the half hour}} \\
\addlinespace
& $\beta_E/\eta$ & 0.022709 & 0.022723 & 0.000014 & 0.000097 & [0.022532, 0.022915] \\
& $\beta_L/\eta$ & 0.053560 & 0.053563 & 0.000003 & 0.000320 & [0.052960, 0.054176] \\
& $\mathrm{LateDummy}/\eta$ & 0.221091 & 0.221394 & 0.000302 & 0.009580 & [0.202434, 0.239620] \\
& $\lambda/\eta$ & 0.000000 & 0.000045 & 0.000045 & 0.000075 & [0.000000, 0.000264] \\
& $\delta/\eta$ & 0.047123 & 0.047058 & -0.000065 & 0.000595 & [0.045894, 0.048194] \\
\bottomrule
\end{tabular}

\vspace{0.5em}
\parbox{0.95\textwidth}{\footnotesize
Notes: The desired CBD arrival times are grouped into 60-minute intervals. Specifications with a late dummy include an additional fixed penalty for arriving after the desired CBD arrival time. For the bootstrap, commuter counts in each station--train cell are rounded to the nearest integer, so the point estimates may differ slightly from those in the main tables. The bootstrap SE is the sample standard deviation of the 1,000 retained estimates. All 1,000 bootstrap re-estimations terminated successfully in each specification. The 95\% intervals are the 2.5th and 97.5th percentiles of those estimates. These quantities condition on the expanded population and do not include survey-sampling uncertainty.
}
\end{table}

\section{Bootstrap simulation diagnostics}
\label{app:bootstrap-diagnostics}

This appendix examines the numerical reliability of the
bootstrap results. Appendix~\ref{app:bootstrap-sensitivity}
compares the bootstrap results across simulation settings,
varying the number of retained states $B$, the sampling
interval, and the burn-in. Appendix~\ref{app:finite-simulation-error}
then examines whether the baseline value of $B=1{,}000$ is
large enough to compute the bootstrap standard errors precisely.
Because the bootstrap standard error in \eqref{eq:bootstrap-se}
is computed from a finite number of states generated by a
Markov chain, it is subject to simulation error
\citep[Section~6.4]{efron1993}. We quantify this error
using the Monte Carlo standard error (MCSE), accounting for
serial correlation in the retained states.
Appendix~\ref{app:seed-robustness} provides a complementary
check by repeating the bootstrap with different random seeds.

\subsection{Sensitivity to simulation settings}
\label{app:bootstrap-sensitivity}

Table~\ref{tab:bootstrap-diagnostics} assesses whether the bootstrap
mean, bias, standard error, and percentile interval are sensitive
to the number of retained states, sampling interval, and burn-in.

\begin{table}[htbp]
\centering
\caption{Sensitivity and finite-simulation precision of the bootstrap}
\label{tab:bootstrap-diagnostics}
\scriptsize
\setlength{\tabcolsep}{3pt}

\begin{tabular}{llrrrr}
\toprule
\multicolumn{6}{l}{\textit{Panel A: Number of retained states}} \\
\addlinespace
$B$ & Parameter & Bootstrap mean & Bias & Bootstrap SE & 95\% interval \\
\midrule
500   & $\beta_E/\eta$ & 0.018030 & -0.000010 & 0.000126 & [0.017806, 0.018272] \\
      & $\beta_L/\eta$ & 0.022453 & -0.000013 & 0.000134 & [0.022205, 0.022722] \\
      & $\lambda/\eta$ & 0.001812 & -0.000036 & 0.000162 & [0.001490, 0.002180] \\
      & $\delta/\eta$  & 0.046312 &  0.000091 & 0.000634 & [0.045021, 0.047515] \\
\addlinespace
1,000 & $\beta_E/\eta$ & 0.018021 & -0.000018 & 0.000128 & [0.017787, 0.018266] \\
      & $\beta_L/\eta$ & 0.022446 & -0.000019 & 0.000136 & [0.022172, 0.022707] \\
      & $\lambda/\eta$ & 0.001803 & -0.000045 & 0.000166 & [0.001480, 0.002169] \\
      & $\delta/\eta$  & 0.046320 &  0.000100 & 0.000647 & [0.045028, 0.047521] \\
\addlinespace
1,500 & $\beta_E/\eta$ & 0.018024 & -0.000016 & 0.000125 & [0.017790, 0.018263] \\
      & $\beta_L/\eta$ & 0.022450 & -0.000016 & 0.000136 & [0.022171, 0.022706] \\
      & $\lambda/\eta$ & 0.001804 & -0.000044 & 0.000166 & [0.001481, 0.002161] \\
      & $\delta/\eta$  & 0.046330 &  0.000110 & 0.000661 & [0.045016, 0.047565] \\
\bottomrule
\end{tabular}

\vspace{1em}

\begin{tabular}{lllrrrr}
\toprule
\multicolumn{7}{l}{\textit{Panel B: Sampling interval and burn-in}} \\
\addlinespace
Interval & Burn-in & Parameter & Bootstrap mean & Bias & Bootstrap SE
         & 95\% interval \\
\midrule
$N$  & 0     & $\beta_E/\eta$ & 0.018024 & -0.000016 & 0.000130 & [0.017789, 0.018270] \\
     &       & $\beta_L/\eta$ & 0.022449 & -0.000016 & 0.000138 & [0.022172, 0.022714] \\
     &       & $\lambda/\eta$ & 0.001807 & -0.000041 & 0.000178 & [0.001480, 0.002171] \\
     &       & $\delta/\eta$  & 0.046311 &  0.000091 & 0.000655 & [0.045028, 0.047521] \\
\addlinespace
$N$  & $10N$ & $\beta_E/\eta$ & 0.018021 & -0.000018 & 0.000128 & [0.017787, 0.018266] \\
     &       & $\beta_L/\eta$ & 0.022446 & -0.000019 & 0.000136 & [0.022172, 0.022707] \\
     &       & $\lambda/\eta$ & 0.001803 & -0.000045 & 0.000166 & [0.001480, 0.002169] \\
     &       & $\delta/\eta$  & 0.046320 &  0.000100 & 0.000647 & [0.045028, 0.047521] \\
\addlinespace
$N$  & $20N$ & $\beta_E/\eta$ & 0.018021 & -0.000019 & 0.000127 & [0.017787, 0.018265] \\
     &       & $\beta_L/\eta$ & 0.022446 & -0.000020 & 0.000136 & [0.022169, 0.022704] \\
     &       & $\lambda/\eta$ & 0.001803 & -0.000045 & 0.000168 & [0.001480, 0.002171] \\
     &       & $\delta/\eta$  & 0.046319 &  0.000099 & 0.000650 & [0.045010, 0.047521] \\
\addlinespace
$2N$ & $10N$ & $\beta_E/\eta$ & 0.018025 & -0.000015 & 0.000125 & [0.017787, 0.018262] \\
     &       & $\beta_L/\eta$ & 0.022447 & -0.000018 & 0.000137 & [0.022175, 0.022707] \\
     &       & $\lambda/\eta$ & 0.001809 & -0.000039 & 0.000168 & [0.001488, 0.002168] \\
     &       & $\delta/\eta$  & 0.046312 &  0.000092 & 0.000709 & [0.044938, 0.047625] \\
\addlinespace
$3N$ & $10N$ & $\beta_E/\eta$ & 0.018024 & -0.000016 & 0.000126 & [0.017778, 0.018267] \\
     &       & $\beta_L/\eta$ & 0.022450 & -0.000015 & 0.000134 & [0.022185, 0.022711] \\
     &       & $\lambda/\eta$ & 0.001810 & -0.000038 & 0.000168 & [0.001503, 0.002168] \\
     &       & $\delta/\eta$  & 0.046325 &  0.000105 & 0.000672 & [0.044996, 0.047578] \\
\bottomrule
\end{tabular}

\vspace{0.5em}
\parbox{0.95\textwidth}{\footnotesize
Notes: Panel A holds the sampling interval at $N$, burn-in at $10N$,
and the random seed at 20260821. Panel B sets $B=1{,}000$ and
uses the same seed. The baseline specification is $B=1{,}000$,
a sampling interval of $N$, and a burn-in of $10N$.
Each nonbaseline setting in Panel B changes only one of the latter
two settings. The 95\% intervals are the 2.5th and 97.5th
percentiles of the retained estimates.
}
\end{table}

Panel A assesses the sensitivity of the four reported bootstrap quantities to
the number of retained states under the baseline sampling interval and burn-in.  The bootstrap mean, bias, standard error, and percentile interval are stable across $B=500, 1{,}000$ and $1{,}500$. Panel B varies the sampling
interval and burn-in one at a time around the baseline.
With a burn-in of $10N$, increasing the sampling interval from $N$ to $2N$
changes the bootstrap SEs by $-2.3\%$ for $\beta_E/\eta$, $+0.7\%$ for
$\beta_L/\eta$, $+1.2\%$ for $\lambda/\eta$, and $+9.6\%$ for
$\delta/\eta$. With a sampling interval of $N$, introducing a burn-in of
$10N$ changes them by $-1.5\%$, $-1.4\%$, $-6.7\%$, and $-1.2\%$,
respectively. Increasing the burn-in from $10N$ to $20N$ or the sampling
interval from $N$ to $3N$ also leaves the results broadly stable. Although the bootstrap SE of $\delta/\eta$ changes by about 10\% in one of these comparisons, the change does not affect our conclusions because the SE is small relative to the point estimate.

\subsection{Finite-simulation error of bootstrap standard errors}
\label{app:finite-simulation-error}

The bootstrap variance $\widehat V_{\mathrm{boot}}
=\widehat{\operatorname{SE}}_{\mathrm{boot}}^2$ is, up to the factor
$B/(B-1)$, the sample mean of the series
$z^{(b)}=(\widehat\theta^{b}-\overline{\widehat\theta})^2$, $b=1,\ldots,B$.
Its simulation error is therefore that of a sample mean, and the MCSE of the
bootstrap standard error follows from the MCSE of this mean by the delta
method \citep[Section~1.13]{geyer2026}.  If the retained estimates were
independent, the variance of the sample mean would be
$\operatorname{Var}(z^{(b)})/B$.  Because the retained states are generated
by a Markov chain, however, successive $z^{(b)}$ are serially correlated,
and $\operatorname{Var}(z^{(b)})$ must be replaced by the long-run variance
$\Omega_z=\gamma_{z,0}+2\sum_{k=1}^{\infty}\gamma_{z,k}$, where
$\gamma_{z,k}$ is the lag-$k$ autocovariance of $z^{(b)}$.  For a
sufficiently long stationary run,
\begin{equation}
    \operatorname{MCSE}
    \left(\widehat{\operatorname{SE}}_{\mathrm{boot}}\right)
    \simeq
    \frac{1}{2\widehat{\operatorname{SE}}_{\mathrm{boot}}}
    \sqrt{\frac{\Omega_z}{B}}.
    \label{eq:bootstrap-se-mcse}
\end{equation}
We estimate $\Omega_z$ by replacing the autocovariances with their sample
counterparts and truncating the sum at a lag $K$.  The truncation lag is
selected by a consistent spectral-variance rule, such as the
initial-positive-sequence estimator, rather than by imposing an AR(1)
approximation on the retained parameter series.

This MCSE can be interpreted through the effective sample size (ESS) of
$z^{(b)}$, $B_{\mathrm{eff},z}=B/(1+2\sum_{k=1}^{\infty}\rho_{z,k})$, where
$\rho_{z,k}$ is the lag-$k$ autocorrelation of $z^{(b)}$.  If
$\widehat\theta^{b}$ is approximately normal,
$\operatorname{Var}(z^{(b)})\simeq 2\widehat V_{\mathrm{boot}}^2$, and
\eqref{eq:bootstrap-se-mcse} reduces to
\begin{equation}
    \frac{
      \operatorname{MCSE}
      (\widehat{\operatorname{SE}}_{\mathrm{boot}})
    }{
      \widehat{\operatorname{SE}}_{\mathrm{boot}}
    }
    \simeq \frac{1}{\sqrt{2B_{\mathrm{eff},z}}}.
    \label{eq:relative-mcse}
\end{equation}
The relevant ESS is therefore that of $z^{(b)}$, not necessarily that of
$\widehat\theta^{b}$ itself.  For example, a relative MCSE of 5\% requires
$B_{\mathrm{eff},z}\approx 200$.

The 5\% relative-MCSE threshold introduced in Section~\ref{subsec:model-based-bootstrap} is a computational-precision criterion.
The MCSE is reported separately and is not added to the bootstrap standard
error.  It also does not detect bias caused by failure to reach the
invariant distribution, which must be assessed with separate convergence and
initialization checks.

Table~\ref{tab:bootstrap-mcse} reports the finite-simulation
precision of the bootstrap standard errors under the baseline
simulation settings.

\begin{table}[htbp]
\centering
\caption{Finite-simulation precision of bootstrap standard errors}
\label{tab:bootstrap-mcse}
\small
\begin{tabular}{lrrr}
\toprule
\addlinespace
Parameter & $B_{\mathrm{eff},z}$
          & $\operatorname{MCSE}(\widehat{\operatorname{SE}}_{\mathrm{boot}})$
          & Relative MCSE \\
\midrule
$\beta_E/\eta$ & 778.476 & 0.000003167 & 2.479\% \\
$\beta_L/\eta$ & 780.234 & 0.000003397 & 2.504\% \\
$\lambda/\eta$ & 809.181 & 0.000004295 & 2.581\% \\
$\delta/\eta$  & 712.252 & 0.000018023 & 2.784\% \\
\bottomrule
\end{tabular}

\vspace{0.5em}
\parbox{0.95\textwidth}{\footnotesize Notes: The results are based on the baseline bootstrap with
$B=1{,}000$, a sampling interval of $N$, and a burn-in of $10N$.
$B_{\mathrm{eff},z}$ denotes the effective sample size of the
squared-centered series. Relative MCSE is the Monte Carlo standard
error of the bootstrap SE divided by the bootstrap SE.}
\end{table}

Under the baseline simulation settings, the relative MCSE is
2.479\% for $\beta_E/\eta$, 2.504\% for
$\beta_L/\eta$, 2.581\% for $\lambda/\eta$, and
2.784\% for $\delta/\eta$. All values are below
the 5\% operational precision target.
Appendix~\ref{app:seed-robustness} provides an additional check
using different random seeds.

\subsection{Random-seed robustness}
\label{app:seed-robustness}

We repeat the baseline bootstrap with three random seeds, 20260821, 20260929, and 20261030, keeping all other settings fixed. If the MCSE is accurate, the bootstrap standard errors should differ across seeds by amounts of the same order as the MCSE. For each seed, we compare the bootstrap mean, bootstrap standard error, and relative MCSE. This exercise complements the finite-simulation precision assessment in Table~\ref{tab:bootstrap-mcse}. Table~\ref{tab:bootstrap-seed-robustness} reports the results for the three random seeds.

\begin{table}[htbp]
\centering
\caption{Random-seed robustness of the baseline bootstrap}
\label{tab:bootstrap-seed-robustness}
\small
\begin{tabular}{llrrr}
\toprule
Seed & Parameter & Bootstrap mean & Bootstrap SE & Relative MCSE \\
\midrule
20260821 & $\beta_E/\eta$ & 0.018021 & 0.000128 & 2.479\% \\
         & $\beta_L/\eta$ & 0.022446 & 0.000136 & 2.504\% \\
         & $\lambda/\eta$ & 0.001803 & 0.000166 & 2.581\% \\
         & $\delta/\eta$  & 0.046320 & 0.000647 & 2.784\% \\
\addlinespace
20260929 & $\beta_E/\eta$ & 0.018015 & 0.000122 & 2.373\% \\
         & $\beta_L/\eta$ & 0.022447 & 0.000136 & 2.394\% \\
         & $\lambda/\eta$ & 0.001812 & 0.000169 & 2.446\% \\
         & $\delta/\eta$  & 0.046304 & 0.000692 & 2.739\% \\
\addlinespace
20261030 & $\beta_E/\eta$ & 0.018023 & 0.000126 & 2.914\% \\
         & $\beta_L/\eta$ & 0.022453 & 0.000144 & 2.854\% \\
         & $\lambda/\eta$ & 0.001824 & 0.000172 & 2.566\% \\
         & $\delta/\eta$  & 0.046290 & 0.000672 & 2.614\% \\
\bottomrule
\end{tabular}
\end{table}

Across the three seeds, the difference between the largest and
smallest bootstrap means is at most 0.1\% of the smallest value for
$\beta_E/\eta$, $\beta_L/\eta$, and $\delta/\eta$, and 1.2\% for
$\lambda/\eta$. The corresponding differences in the bootstrap
standard errors are 4.9\%, 5.9\%, 3.6\%, and 7.0\% for
$\beta_E/\eta$, $\beta_L/\eta$, $\lambda/\eta$, and $\delta/\eta$,
respectively. The relative MCSE ranges from 2.4\% to 2.9\%
across all parameters and seeds, below the 5\% operational
target in all cases. The reported bootstrap results are therefore
not sensitive to the random-number seed.

\section{Expansion weights and survey sample sizes}
\label{app:survey-ess}

\subsection{Inference conditional on expansion weights}
\label{app:expansion-weight-inference}

As described in Section~\ref{subsec:expansion}, the commuter population at each
station is constructed from the expansion weights of the underlying
travel-survey records, and it is allocated across trains in proportion to the
numbers of survey respondents who choose each train. Both the likelihood and the bootstrap use these expanded
counts: the resulting $N=140{,}473$ commuters form the finite population of the
equilibrium game, and each expanded commuter receives revision opportunities
in the stochastic adjustment process.

This treatment is appropriate for measuring equilibrium variation conditional
on the expanded population, but the expanded commuters must not be interpreted
as independent survey respondents.  Doing so would overstate the precision
of the survey data and could therefore understate, rather than overstate, the
sampling uncertainty associated with the survey and the construction of the
weights.  The bootstrap SE is consequently not a complete survey
sampling SE: it conditions on the expansion weights and the expanded
station populations and excludes variation from sampling, nonresponse, and
recalculation of the weights.

Because the bootstrap does not capture this uncertainty, the rest of this
appendix examines it in two ways. Appendix~\ref{app:station-survey-ess}
reports the number of survey respondents and the effective sample size at
each station, which indicate how precisely the train shares within each
station are measured. Appendix~\ref{app:low-ess-sensitivity} then examines
whether the estimates are sensitive to the train allocations at the stations
with the fewest respondents, where this uncertainty is largest.

\subsection{Survey respondents and effective sample sizes by station}
\label{app:station-survey-ess}

Because the train shares within each station are computed from unweighted
respondent counts (Section~\ref{subsec:expansion}), the precision of the shares at
station $\ell$ depends on its number of respondents,
$m_\ell=\sum_r m_{\ell r}$, where $m_{\ell r}$ denotes the number of
respondents at station $\ell$ who take train $r$.
Table~\ref{tab:survey-ess-by-station} reports $m_\ell$ together with the
expanded station population.  Within the estimation window, the respondent
counts sum to 2,740 and the expanded counts sum to 140,473.  Hon-Kawagoe and
Nakai have the fewest respondents, 14 and 15, respectively.

Table~\ref{tab:survey-ess-by-station} also reports a Kish-type effective
sample size (ESS), which measures the precision of the train shares if they
were instead weighted by the expansion factors, in terms of an equivalent
number of unweighted respondents.  Let
$q_{\ell r}$ denote the sum of the original expansion weights of the
respondents at station $\ell$ who take train $r$.  We approximate the
individual weights in each cell by the cell-average weight
$q_{\ell r}/m_{\ell r}$.\footnote{This approximation assigns the same average weight
to every respondent in a station--train cell.
It therefore cannot distinguish equal weights from
a case in which one or a few respondents with
exceptionally large weights account for most
of the expanded count.
The survey ESS should therefore be recalculated directly
from the respondent-level weights, rather than from
the cell averages, so that this within-cell weight
dispersion is reflected in the diagnostic.}  The
resulting station-specific ESS is
\begin{equation}
    \widehat n_{\mathrm{eff},\ell}^{S}
    =
    \frac{\left(\sum_r q_{\ell r}\right)^2}
         {\sum_r q_{\ell r}^2/m_{\ell r}}.
    \label{eq:station-survey-ess}
\end{equation}
The station-level ESS values sum to 1,699.2, or 62.0\% of the 2,740
respondents, and the ratio falls to 37.9\% at Hon-Kawagoe.  Weighting the
train shares by the expansion factors would therefore substantially reduce
their precision, particularly at stations with few respondents.  The
unweighted shares used in this study avoid this loss.  In exchange, they assume that train choices do not differ
systematically across the groups that determine the expansion factors, as
discussed in Section~\ref{subsec:expansion}.

\begin{table}[p]
\centering
\caption{Survey respondents and effective sample sizes by station}
\label{tab:survey-ess-by-station}
\scriptsize
\setlength{\tabcolsep}{5pt}
\begin{tabular}{rlrrrr}
\toprule
ID & Station & Survey count $m_\ell$ & Expanded count $q_\ell$
   & ESS & ESS/$m_\ell$ \\
\midrule
1  & Hon-Kawagoe      & 14  & 1,042  & 5.3   & 37.9\% \\
2  & Minami-Otsuka    & 23  & 1,057  & 19.0  & 82.5\% \\
3  & Shin-Sayama      & 27  & 966    & 18.8  & 69.5\% \\
4  & Sayamashi        & 105 & 5,413  & 75.7  & 72.1\% \\
5  & Iriso            & 28  & 651    & 23.9  & 85.5\% \\
6  & Shin-Tokorozawa  & 140 & 4,493  & 112.0 & 80.0\% \\
7  & Kokukoen         & 32  & 1,193  & 21.7  & 67.8\% \\
8  & Tokorozawa       & 78  & 2,622  & 48.9  & 62.7\% \\
9  & Higashi-Murayama & 109 & 6,171  & 60.4  & 55.4\% \\
10 & Kumegawa         & 114 & 5,789  & 57.4  & 50.4\% \\
11 & Kodaira          & 460 & 26,058 & 333.9 & 72.6\% \\
12 & Hana-Koganei     & 253 & 10,412 & 151.4 & 59.8\% \\
13 & Tanashi          & 301 & 11,664 & 212.4 & 70.6\% \\
14 & Seibu-Yagisawa   & 77  & 4,507  & 34.0  & 44.2\% \\
15 & Higashi-Fushimi  & 85  & 5,061  & 41.9  & 49.3\% \\
16 & Musashiseki      & 118 & 7,908  & 76.2  & 64.6\% \\
17 & Kami-Shakujii    & 205 & 7,829  & 134.0 & 65.4\% \\
18 & Kami-Igusa       & 52  & 2,219  & 27.4  & 52.6\% \\
19 & Iogi             & 70  & 4,019  & 39.1  & 55.9\% \\
20 & Shimo-Igusa      & 57  & 4,385  & 21.6  & 37.9\% \\
21 & Saginomiya       & 119 & 7,393  & 48.2  & 40.5\% \\
22 & Toritsu-Kasei    & 59  & 4,157  & 21.0  & 35.5\% \\
23 & Nogata           & 67  & 5,377  & 29.3  & 43.7\% \\
24 & Numabukuro       & 67  & 4,598  & 47.6  & 71.1\% \\
25 & Araiyakushi-mae  & 36  & 2,780  & 19.9  & 55.4\% \\
26 & Nakai            & 15  & 1,464  & 7.1   & 47.1\% \\
27 & Shimo-Ochiai     & 29  & 1,242  & 11.0  & 37.9\% \\
\midrule
\multicolumn{2}{l}{Total} & 2,740 & 140,473 & 1,699.2 & 62.0\% \\
\bottomrule
\end{tabular}

\vspace{0.5em}
\parbox{0.95\textwidth}{\footnotesize Notes: Expanded counts are based on the
sample-share allocation of station-level expansion totals used in the
estimation. ESS is calculated using the original expansion weights before
sample-share allocation. This simplified diagnostic assigns a common
average weight within each station--train cell. If individual weights vary
within a cell, the reported ESS is an upper bound on the individual-level
Kish ESS. The calculation reflects weight dispersion but not clustering,
nonresponse, calibration uncertainty, or the dependence created by
respondents who purchased tickets at two locations.}
\end{table}

\subsection{Sensitivity at stations with few respondents}
\label{app:low-ess-sensitivity}

To examine whether the stations with the fewest survey respondents, Hon-Kawagoe
and Nakai, drive the estimates, we perturb their observed train allocations.  At
each of these stations, we change the number of commuters on the
largest-share train, that is, the train chosen by the most commuters from
that station, by $\pm10\%$ and $\pm20\%$.  The station population is held
fixed by reallocating the difference across the other trains available at
that station in proportion to their observed counts, and the model is then
re-estimated.  Table~\ref{tab:low-ess-sensitivity} reports the resulting
estimates. Across all perturbations, the estimates change by at most
0.1\% for $\beta_E/\eta$ and $\beta_L/\eta$, 1.4\% for $\lambda/\eta$, and
0.4\% for $\delta/\eta$ relative to the baseline, and each change is smaller
than the corresponding bootstrap standard error. The estimates are therefore
not driven by the observed train allocations at these two stations.

\begin{table}[p]
\centering
\caption{Sensitivity to train allocations at stations with few survey respondents}
\label{tab:low-ess-sensitivity}
\small
\setlength{\tabcolsep}{6pt}
\begin{tabular}{llrrrr}
\toprule
Station & Perturbation
        & $\beta_E/\eta$ & $\beta_L/\eta$
        & $\lambda/\eta$ & $\delta/\eta$ \\
\midrule
Baseline & Observed
         & 0.018039 & 0.022466 & 0.001848 & 0.046220 \\
         & 
         & (0.000128) & (0.000136) & (0.000166) & (0.000647) \\
\addlinespace
Hon-Kawagoe & $-20\%$
            & 0.018040 & 0.022473 & 0.001868 & 0.046048 \\
            &
            & (0.000128) & (0.000135) & (0.000166) & (0.000648) \\
            & $-10\%$
            & 0.018040 & 0.022470 & 0.001859 & 0.046133 \\
            &
            & (0.000128) & (0.000136) & (0.000166) & (0.000647) \\
            & $+10\%$
            & 0.018038 & 0.022461 & 0.001836 & 0.046310 \\
            &
            & (0.000128) & (0.000135) & (0.000166) & (0.000645) \\
            & $+20\%$
            & 0.018036 & 0.022455 & 0.001823 & 0.046403 \\
            &
            & (0.000127) & (0.000135) & (0.000165) & (0.000645) \\
\addlinespace
Nakai       & $-20\%$
            & 0.018025 & 0.022442 & 0.001836 & 0.046221 \\
            &
            & (0.000127) & (0.000135) & (0.000165) & (0.000644) \\
            & $-10\%$
            & 0.018031 & 0.022455 & 0.001842 & 0.046219 \\
            &
            & (0.000128) & (0.000136) & (0.000166) & (0.000647) \\
            & $+10\%$
            & 0.018046 & 0.022478 & 0.001854 & 0.046222 \\
            &
            & (0.000128) & (0.000136) & (0.000167) & (0.000649) \\
            & $+20\%$
            & 0.018053 & 0.022488 & 0.001859 & 0.046223 \\
            &
            & (0.000128) & (0.000136) & (0.000167) & (0.000647) \\
\bottomrule
\end{tabular}

\vspace{0.5em}
\parbox{0.95\textwidth}{\footnotesize Notes: The perturbation changes the number assigned to the station's largest-share train while holding the station population fixed and reallocating the remainder across other trains in proportion to their observed counts. Point estimates are reported in the first row for each specification, with bootstrap standard errors in parentheses. As in the bootstrap, commuter counts in each station--train cell are rounded to the nearest integer, so the baseline estimates may differ slightly from those in the main tables. Counts and all inputs outside the perturbed station, including exogenous passenger counts, train capacities, and the timetable, are fixed.}
\end{table}

\section{Unobserved cells and station aggregation}
\label{app:unobserved-cells}

The baseline specification treats available station--train cells that are
absent from the commuter-choice data as having zero observed commuters.  This
appendix reports how often such cells occur and examines the sensitivity of the
estimates to this treatment.

Table~\ref{tab:unobserved-cells-by-station} reports the number of available
station--train cells, cells with positive expanded commuter counts, and
unobserved cells at each origin station.  The unobserved-cell share is the
number of unobserved cells divided by the number of available cells.  The incidence varies substantially across stations, from 1.5\% at Kodaira to 74.1\% at Hon-Kawagoe.

\begin{table}[p]
\centering
\caption{Station-specific incidence of unobserved station--train cells}
\label{tab:unobserved-cells-by-station}
\scriptsize
\setlength{\tabcolsep}{5pt}
\begin{tabular}{rlrrrr}
\toprule
ID & Station & Available cells & Positive cells & Unobserved cells
   & Unobserved share \\
\midrule
1  & Hon-Kawagoe      & 27 &  7 & 20 & 74.1\% \\
2  & Minami-Otsuka    & 25 & 10 & 15 & 60.0\% \\
3  & Shin-Sayama      & 25 & 14 & 11 & 44.0\% \\
4  & Sayamashi        & 29 & 24 &  5 & 17.2\% \\
5  & Iriso            & 27 & 15 & 12 & 44.4\% \\
6  & Shin-Tokorozawa  & 43 & 35 &  8 & 18.6\% \\
7  & Kokukoen         & 41 & 19 & 22 & 53.7\% \\
8  & Tokorozawa       & 43 & 30 & 13 & 30.2\% \\
9  & Higashi-Murayama & 43 & 33 & 10 & 23.3\% \\
10 & Kumegawa         & 41 & 36 &  5 & 12.2\% \\
11 & Kodaira          & 65 & 64 &  1 &  1.5\% \\
12 & Hana-Koganei     & 65 & 54 & 11 & 16.9\% \\
13 & Tanashi          & 77 & 66 & 11 & 14.3\% \\
14 & Seibu-Yagisawa   & 48 & 29 & 19 & 39.6\% \\
15 & Higashi-Fushimi  & 48 & 35 & 13 & 27.1\% \\
16 & Musashiseki      & 48 & 36 & 12 & 25.0\% \\
17 & Kami-Shakujii    & 89 & 65 & 24 & 27.0\% \\
18 & Kami-Igusa       & 42 & 27 & 15 & 35.7\% \\
19 & Iogi             & 42 & 25 & 17 & 40.5\% \\
20 & Shimo-Igusa      & 42 & 25 & 17 & 40.5\% \\
21 & Saginomiya       & 89 & 57 & 32 & 36.0\% \\
22 & Toritsu-Kasei    & 42 & 22 & 20 & 47.6\% \\
23 & Nogata           & 42 & 27 & 15 & 35.7\% \\
24 & Numabukuro       & 42 & 30 & 12 & 28.6\% \\
25 & Araiyakushi-mae  & 42 & 23 & 19 & 45.2\% \\
26 & Nakai            & 42 & 11 & 31 & 73.8\% \\
27 & Shimo-Ochiai     & 42 & 18 & 24 & 57.1\% \\
\midrule
\multicolumn{2}{l}{Total} & 1,251 & 837 & 414 & 33.1\% \\
\bottomrule
\end{tabular}

\vspace{0.5em}
\parbox{0.95\textwidth}{\footnotesize Notes: Available cells are
station--train pairs at which the train stops according to the timetable. Positive
cells appear in the commuter-choice data with a positive expanded count.
Unobserved cells are available pairs with no observed commuters. Accordingly,
``unobserved'' does not by itself distinguish true zero demand, a zero in the
underlying survey sample, and a train-matching or data-construction failure.}
\end{table}

To assess the sensitivity of the estimates to the treatment of these cells as
zeros, we construct
alternative zoning schemes by pooling adjacent stations only when they have
identical train choice sets.  Pooling reduces the overall share of unobserved
cells from 33.1\% in the 27-station baseline to
32.9\%, 26.5\%, and 23.2\% in the 24-, 19-,
and 15-zone specifications, respectively.
The maximum unobserved-cell share nevertheless
remains 74.1\% in all three alternative
specifications, so the alternative aggregations
do not eliminate sparsity at every origin.

\begin{table}[p]
\centering
\caption{Station aggregation and unobserved-cell shares}
\label{tab:station-aggregation-statistics}
\scriptsize
\setlength{\tabcolsep}{4pt}
\renewcommand{\arraystretch}{1.0}

\resizebox{\textwidth}{!}{%
\begin{tabular}{llllrrrr}
\toprule
\multicolumn{4}{c}{Stations and zones}
& \multicolumn{4}{c}{Unobserved-cell share} \\
\cmidrule(lr){1-4}
\cmidrule(lr){5-8}
Baseline & 24-zone & 19-zone & 15-zone
& Baseline & 24-zone & 19-zone & 15-zone \\
\midrule
Hon-Kawagoe & Hon-Kawagoe & Hon-Kawagoe & Hon-Kawagoe
& 74.1\% & 74.1\% & 74.1\% & 74.1\% \\
Minami-Otsuka & Minami-Otsuka--Shin-Sayama & Minami-Otsuka--Shin-Sayama & Minami-Otsuka--Shin-Sayama
& 60.0\% & 32.0\% & 32.0\% & 32.0\% \\
Shin-Sayama & & &
& 44.0\% & & & \\
Sayamashi & Sayamashi & Sayamashi & Sayamashi
& 17.2\% & 17.2\% & 17.2\% & 17.2\% \\
Iriso & Iriso & Iriso & Iriso
& 44.4\% & 44.4\% & 44.4\% & 44.4\% \\
Shin-Tokorozawa & Shin-Tokorozawa & Shin-Tokorozawa & Shin-Tokorozawa
& 18.6\% & 18.6\% & 18.6\% & 18.6\% \\
Kokukoen & Kokukoen & Kokukoen & Kokukoen
& 53.7\% & 53.7\% & 53.7\% & 53.7\% \\
Tokorozawa & Tokorozawa--Higashi-Murayama & Tokorozawa--Higashi-Murayama & Tokorozawa--Higashi-Murayama
& 30.2\% & 14.0\% & 14.0\% & 14.0\% \\
Higashi-Murayama & & &
& 23.3\% & & & \\
Kumegawa & Kumegawa & Kumegawa & Kumegawa
& 12.2\% & 12.2\% & 12.2\% & 12.2\% \\
Kodaira & Kodaira--Hana-Koganei & Kodaira--Hana-Koganei & Kodaira--Hana-Koganei
& 1.5\% & 1.5\% & 1.5\% & 1.5\% \\
Hana-Koganei & & &
& 16.9\% & & & \\
Tanashi & Tanashi & Tanashi & Tanashi
& 14.3\% & 14.3\% & 14.3\% & 14.3\% \\
Seibu-Yagisawa & Seibu-Yagisawa & Seibu-Yagisawa--Higashi-Fushimi & Seibu-Yagisawa--Musashiseki
& 39.6\% & 39.6\% & 18.8\% & 8.3\% \\
Higashi-Fushimi & Higashi-Fushimi & &
& 27.1\% & 27.1\% & & \\
Musashiseki & Musashiseki & Musashiseki &
& 25.0\% & 25.0\% & 25.0\% & \\
Kami-Shakujii & Kami-Shakujii & Kami-Shakujii & Kami-Shakujii
& 27.0\% & 27.0\% & 27.0\% & 27.0\% \\
Kami-Igusa & Kami-Igusa & Kami-Igusa--Iogi & Kami-Igusa--Shimo-Igusa
& 35.7\% & 35.7\% & 23.8\% & 21.4\% \\
Iogi & Iogi & &
& 40.5\% & 40.5\% & & \\
Shimo-Igusa & Shimo-Igusa & Shimo-Igusa &
& 40.5\% & 40.5\% & 40.5\% & \\
Saginomiya & Saginomiya & Saginomiya & Saginomiya
& 36.0\% & 36.0\% & 36.0\% & 36.0\% \\
Toritsu-Kasei & Toritsu-Kasei & Toritsu-Kasei--Nogata & Toritsu-Kasei--Shimo-Ochiai
& 47.6\% & 47.6\% & 28.6\% & 4.8\% \\
Nogata & Nogata & &
& 35.7\% & 35.7\% & & \\
Numabukuro & Numabukuro & Numabukuro--Araiyakushi-mae &
& 28.6\% & 28.6\% & 16.7\% & \\
Araiyakushi-mae & Araiyakushi-mae & &
& 45.2\% & 45.2\% & & \\
Nakai & Nakai & Nakai--Shimo-Ochiai &
& 73.8\% & 73.8\% & 42.9\% & \\
Shimo-Ochiai & Shimo-Ochiai & &
& 57.1\% & 57.1\% & & \\
\midrule
\multicolumn{4}{l}{Overall share}
& 33.1\% & 32.9\% & 26.5\% & 23.2\% \\
\bottomrule
\end{tabular}
}
\end{table}

Table~\ref{tab:unobserved-cell-sensitivity} reports the
estimated parameters under the alternative station
aggregations and their percentage changes relative
to the baseline specification.

\begin{table}[htbp]
\centering
\caption{Sensitivity to alternative station aggregations}
\label{tab:unobserved-cell-sensitivity}
\small
\setlength{\tabcolsep}{6pt}
\resizebox{\textwidth}{!}{%
\begin{tabular}{lrrrr}
\toprule
Specification & $\beta_E/\eta$ (change) & $\beta_L/\eta$ (change)
              & $\lambda/\eta$ (change) & $\delta/\eta$ (change) \\
\midrule
27-station (Baseline) & 0.018042 & 0.022468 & 0.001845 & 0.046225 \\
24-zone & 0.018050 ($+0.05\%$) & 0.022473 ($+0.02\%$)
        & 0.001829 ($-0.90\%$) & 0.046575 ($+0.76\%$) \\
19-zone & 0.018068 ($+0.15\%$) & 0.022493 ($+0.11\%$)
        & 0.001873 ($+1.53\%$) & 0.046419 ($+0.42\%$) \\
15-zone & 0.018112 ($+0.39\%$) & 0.022540 ($+0.32\%$)
        & 0.001958 ($+6.09\%$) & 0.046346 ($+0.26\%$) \\
\bottomrule
\end{tabular}
}

\vspace{0.5em}
\parbox{0.95\textwidth}{\footnotesize Notes: Alternative zoning schemes aggregate adjacent stations. Changes in parentheses are relative to the baseline point estimates for the 27-station specification.}
\end{table}

The point estimates remain stable across these specifications.  Pooling mechanically converts
a cell into a positive cell whenever the corresponding train is observed at
any constituent station and also changes the spatial aggregation of the
model.  The exercise should therefore be interpreted as an indirect
sensitivity analysis of the baseline treatment of unobserved cells as zeros,
rather than as an imputation of unobserved commuter counts.  Subject to this
qualification, the results indicate that the main point estimates are not
driven by the station-level incidence of unobserved cells.

\clearpage
\section{Potential-game condition}
\label{app:potential-game-condition}

This appendix shows that $\Phi_N$ in \eqref{eq:finite-potential} satisfies the potential-game condition \eqref{eq:potential-difference}: when a single commuter switches trains, the change in $\Phi_N$ equals the negative of the change in that commuter's cost. Fix $\ell\in[L]$, $r,s\in\mathcal R^\ell$ with $r\neq s$, and $\bm x\in\mathcal X_N$ with $x_r^\ell>0$, and let $\bm x'=\bm x+\tfrac1N(\bm e_s^\ell-\bm e_r^\ell)$. For each train $q$ and segment $k$, let
\begin{equation}
X_q^k(\bm x)=Ny_q^k(\bm x)
\end{equation}
denote the number of endogenous commuters on train $q$ in segment $k$. The switch from $\bm x$ to $\bm x'$ changes only these counts for trains $r$ and $s$ in segments $k\ge\ell$: $X_s^k(\bm x')=X_s^k(\bm x)+1$ and $X_r^k(\bm x')=X_r^k(\bm x)-1$ for all $k\ge\ell$, while all other counts are unchanged.

The first term of \eqref{eq:finite-potential} changes by
\begin{equation}
-\left[\delta T_s^\ell+\beta_E(t^\ast-t_s)_++\beta_L(t_s-t^\ast)_+\right]
+\left[\delta T_r^\ell+\beta_E(t^\ast-t_r)_++\beta_L(t_r-t^\ast)_+\right].
\end{equation}
In the second term, for each segment $k\ge\ell$, the inner sum for train $s$ gains the summand with $n=X_s^k(\bm x')$, and the inner sum for train $r$ loses the summand with $n=X_r^k(\bm x)$. The second term therefore changes by
\begin{equation}
-\sum_{k=\ell}^L\tau_s^k g\left(\frac{E_s^k+X_s^k(\bm x')}{K_s}\right)
+\sum_{k=\ell}^L\tau_r^k g\left(\frac{E_r^k+X_r^k(\bm x)}{K_r}\right)
=-G_s^\ell(\bm x')+G_r^\ell(\bm x).
\end{equation}
Adding the two changes yields $\Phi_N(\bm x')-\Phi_N(\bm x)=-[C_s^\ell(\bm x')-C_r^\ell(\bm x)]$, which is \eqref{eq:potential-difference}. The case $r=s$ is trivial.

In the large-population limit, the same structure holds for the limit
potential $\phi$ in \eqref{eq:limit-potential}.  Differentiating
\eqref{eq:limit-potential} gives
$\partial\phi/\partial x_r^\ell=-C_r^\ell(\bm x)$ for all $\ell\in[L]$ and
$r\in\mathcal R^\ell$.  That is, the gradient of $\phi$ equals the negative
of the cost vector, so $\phi$ is an exact potential of the game with a
continuum of agents, the counterpart of \eqref{eq:potential-difference}.
For continuously differentiable costs, such a potential exists if and only
if the Jacobian matrix of the cost functions is symmetric, a condition that
\citet[Chapter~3]{sandholm2010} calls externality symmetry:
\begin{equation}
\forall \ell, m\in [L], \forall r\in \mathcal{R}^\ell, \forall s\in \mathcal{R}^m, \frac{\partial C_r^\ell}{\partial x_s^m} = \frac{\partial C_s^m}{\partial x_r^\ell}.
\end{equation}
This symmetry can also be verified directly.  If the trains differ, both
sides are zero because there is no externality across trains.  If the
trains are the same, i.e., $r\in\mathcal{R}^\ell\cap\mathcal{R}^m$,
\begin{equation}
\frac{\partial C_r^\ell}{\partial x_r^m} = \sum_{k=\max\{\ell,m\}}^L \frac{N\tau_r^k}{K_r} g'\left(\frac{E_r^k+Ny_r^k(\bm x)}{K_r}\right),
\end{equation}
which is symmetric in $\ell$ and $m$.

\printbibliography

@book{sandholm2010,
  title = {Population Games and Evolutionary Dynamics},
  author = {Sandholm, William H.},
  year = 2010,
  series = {Economic Learning and Social Evolution},
  publisher = {MIT Press},
  address = {Cambridge, Mass},
  isbn = {978-0-262-19587-4}
}

@book{bubeck2015,
  title = {Convex Optimization: Algorithms and Complexity},
  shorttitle = {Convex Optimization},
  author = {Bubeck, S{\'e}bastien},
  year = 2015,
  series = {Foundations and Trends in Machine Learning},
  number = {Volume 8, Issue 3-4},
  publisher = {Now},
  address = {Boston Delft},
  isbn = {978-1-60198-860-7},
  langid = {english}
}

@article{mele2017Econometrica,
  title = {A {{Structural Model}} of {{Dense Network Formation}}},
  author = {Mele, A.},
  year = 2017,
  month = may,
  journal = {Econometrica},
  volume = {85},
  number = {3},
  pages = {825--850},
  issn = {0012-9682},
  doi = {10.3982/ecta10400},
  langid = {english}
}

@article{depalma2015TransportationResearchPartB:Methodological,
  title = {Discomfort in mass transit and its implication for scheduling and pricing},
  volume = {71},
  issn = {01912615},
  url = {https://linkinghub.elsevier.com/retrieve/pii/S0191261514001702},
  doi = {10.1016/j.trb.2014.10.001},
  pages = {1--18},
  journaltitle = {Transportation Research Part B: Methodological},
  shortjournal = {Transportation Research Part B: Methodological},
  author = {De Palma, André and Kilani, Moez and Proost, Stef},
  urldate = {2025-02-19},
  date = {2015-01},
  langid = {english}
}

@article{depalma2017JournalofUrbanEconomics,
  title = {The economics of crowding in rail transit},
  volume = {101},
  issn = {00941190},
  url = {https://linkinghub.elsevier.com/retrieve/pii/S0094119017300566},
  doi = {10.1016/j.jue.2017.06.003},
  pages = {106--122},
  journaltitle = {Journal of Urban Economics},
  shortjournal = {Journal of Urban Economics},
  author = {De Palma, André and Lindsey, Robin and Monchambert, Guillaume},
  urldate = {2025-02-19},
  date = {2017-09},
  langid = {english}
}

@article{kreindler2024ECTA,
  title = {Peak‐Hour Road Congestion Pricing: Experimental Evidence and Equilibrium Implications},
  volume = {92},
  issn = {0012-9682},
  url = {https://www.econometricsociety.org/doi/10.3982/ECTA18422},
  doi = {10.3982/ECTA18422},
  shorttitle = {Peak‐Hour Road Congestion Pricing},
  pages = {1233--1268},
  number = {4},
  journaltitle = {Econometrica},
  shortjournal = {{ECTA}},
  author = {Kreindler, Gabriel},
  urldate = {2025-02-18},
  date = {2024},
  langid = {english}
}

@article{xu2024TransportationResearchPartB:Methodological,
  title = {Estimation of schedule preference and crowding perception in urban rail corridor commuting: An inverse optimization method},
  volume = {189},
  issn = {01912615},
  url = {https://linkinghub.elsevier.com/retrieve/pii/S0191261524001474},
  doi = {10.1016/j.trb.2024.103023},
  shorttitle = {Estimation of schedule preference and crowding perception in urban rail corridor commuting},
  pages = {103023},
  journaltitle = {Transportation Research Part B: Methodological},
  shortjournal = {Transportation Research Part B: Methodological},
  author = {Xu, Pu and Liu, Tian-Liang and Tian, Qiong and Si, Bingfeng and Liu, Wei and Huang, Hai-Jun},
  urldate = {2026-04-09},
  date = {2024-11},
  langid = {english}
}

@incollection{geyer2026,
  title = {Introduction to {{MCMC}}},
  booktitle = {Handbook of Markov Chain Monte Carlo},
  author = {Geyer, Charles J.},
  editor = {Craiu, Radu V. and Vats, Dootika and Jones, Galin L. and Brooks, Steve and Gelman, Andrew and Meng, Xiao-Li},
  year = {2026},
  month = mar,
  edition = {2},
  pages = {1--46},
  publisher = {{Chapman and Hall/CRC}},
  address = {New York},
  doi = {10.1201/9781003453420-1},
  chapter = {1}
}

@article{monderer1996GamesEcon.Behav.,
  title = {Potential Games},
  author = {Monderer, D. and Shapley, L. S.},
  year = 1996,
  month = may,
  journal = {Games and Economic Behavior},
  volume = {14},
  number = {1},
  pages = {124--143},
  issn = {0899-8256},
  doi = {10.1006/game.1996.0044}
}

@article{rosenthal1973IntJGameTheory,
  title = {A Class of Games Possessing Pure-Strategy {{Nash}} Equilibria},
  author = {Rosenthal, Robert W.},
  year = 1973,
  month = dec,
  journal = {International Journal of Game Theory},
  volume = {2},
  number = {1},
  pages = {65--67},
  issn = {0020-7276, 1432-1270},
  doi = {10.1007/BF01737559},
  urldate = {2026-10-05},
  langid = {english}
}

@book{boyd2004,
  title = {Convex Optimization},
  author = {Boyd, Stephen P. and Vandenberghe, Lieven},
  year = 2004,
  publisher = {Cambridge University Press},
  address = {Cambridge, UK ; New York},
  isbn = {978-0-521-83378-3}
}

@article{li2011TransportPolicy,
  title = {Crowding and Public Transport: {{A}} Review of Willingness to Pay Evidence and Its Relevance in Project Appraisal},
  shorttitle = {Crowding and Public Transport},
  author = {Li, Zheng and Hensher, David A.},
  year = 2011,
  month = nov,
  journal = {Transport Policy},
  volume = {18},
  number = {6},
  pages = {880--887},
  issn = {0967070X},
  urldate = {2026-10-15},
  langid = {english}
}

@article{wardman2011TransportReviews,
  title = {Twenty {{Years}} of {{Rail Crowding Valuation Studies}}: {{Evidence}} and {{Lessons}} from {{British Experience}}},
  shorttitle = {Twenty {{Years}} of {{Rail Crowding Valuation Studies}}},
  author = {Wardman, Mark and Whelan, Gerard},
  year = 2011,
  month = may,
  journal = {Transport Reviews},
  volume = {31},
  number = {3},
  pages = {379--398},
  issn = {0144-1647, 1464-5327},
  urldate = {2026-10-15},
  langid = {english}
}

@book{internationaltransportforum2014,
  title = {Valuing {{Convenience}} in {{Public Transport}}},
  author = {{International Transport Forum}},
  year = 2014,
  month = dec,
  series = {{{ITF Round Tables}}},
  publisher = {OECD Publishing},
  address = {Paris},
  isbn = {978-92-821-0767-6},
  langid = {english},
  shortauthor = {ITF}
}

@book{internationaltransportforum2021,
  title = {Decongesting Our {{Cities}}: {{Summary}} and {{Conclusions}}},
  shorttitle = {Decongesting Our {{Cities}}},
  author = {{International Transport Forum}},
  year = 2021,
  month = may,
  series = {{{ITF Roundtable Reports}}},
  number = {183},
  publisher = {OECD Publishing},
  address = {Paris},
  isbn = {978-92-821-3414-6},
  langid = {english},
  shortauthor = {ITF}
}

@article{tirachini2016TransportPolicy,
  title = {Valuation of Sitting and Standing in Metro Trains Using Revealed Preferences},
  author = {Tirachini, Alejandro and Sun, Lijun and Erath, Alexander and Chakirov, Artem},
  year = 2016,
  month = apr,
  journal = {Transport Policy},
  volume = {47},
  pages = {94--104},
  issn = {0967070X},
  doi = {10.1016/j.tranpol.2015.12.004},
  urldate = {2026-10-15},
  langid = {english}
}

@article{horcher2017TransportationResearchPartB:Methodological,
  title = {Crowding Cost Estimation with Large Scale Smart Card and Vehicle Location Data},
  author = {H{\"o}rcher, Daniel and Graham, Daniel J. and Anderson, Richard J.},
  year = 2017,
  month = jan,
  journal = {Transportation Research Part B: Methodological},
  volume = {95},
  pages = {105--125},
  issn = {01912615},
  doi = {10.1016/j.trb.2016.10.015},
  urldate = {2026-10-15},
  langid = {english}
}

@article{peer2016TransportationResearchPartB:Methodological,
  title = {Train Commuters' Scheduling Preferences: {{Evidence}} from a Large-Scale Peak Avoidance Experiment},
  shorttitle = {Train Commuters' Scheduling Preferences},
  author = {Peer, Stefanie and Knockaert, Jasper and Verhoef, Erik T.},
  year = 2016,
  month = jan,
  journal = {Transportation Research Part B: Methodological},
  volume = {83},
  pages = {314--333},
  issn = {01912615},
  doi = {10.1016/j.trb.2015.11.017},
  urldate = {2026-10-15},
  langid = {english}
}

@article{kobayashi2020J.JSCESer.D3,
  title = {Economic Evaluation for Delay and Punctuality of Urban Rail Transit Based on Scheduling Approach},
  author = {Kobayashi, W. and Fukuda, D. and Iwakura, S.},
  year = 2020,
  journal = {Journal of Japan Society of Civil Engineers, Ser. D3 (Infrastructure Planning and Management)},
  volume = {76},
  number = {3},
  pages = {236--250}
}

@article{small1982Am.Econ.Rev.,
  title = {The Scheduling of Consumer Activities: {{Work}} Trips},
  author = {Small, Kenneth A.},
  year = 1982,
  journal = {The American Economic Review},
  volume = {72},
  number = {3},
  pages = {467--479}
}

@article{yap2020TransportmetricaA:TransportScience,
  title = {Crowding Valuation in Urban Tram and Bus Transportation Based on Smart Card Data},
  author = {Yap, Menno and Cats, Oded and Van Arem, Bart},
  year = 2020,
  month = dec,
  journal = {Transportmetrica A: Transport Science},
  volume = {16},
  number = {1},
  pages = {23--42},
  issn = {2324-9935, 2324-9943},
  doi = {10.1080/23249935.2018.1537319},
  urldate = {2026-10-15},
  langid = {english}
}

@techreport{fujishima2025RIETIDP,
  type = {{{RIETI Discussion Paper}}},
  title = {Quantifying Congestion Externalities in Road Networks: {{A}} Structural Estimation Approach Using Stochastic Evolutionary Model},
  author = {Fujishima, Shota and Sakai, Takara and Takayama, Yuki},
  year = 2025,
  month = jun,
  number = {25-E-055},
  address = {Tokyo},
  institution = {{Research Institute of Economy, Trade and Industry}}
}

@article{arnott1990JournalofUrbanEconomics,
  title = {Economics of a Bottleneck},
  author = {Arnott, Richard and De Palma, Andr{\'e} and Lindsey, Robin},
  year = 1990,
  month = jan,
  journal = {Journal of Urban Economics},
  volume = {27},
  number = {1},
  pages = {111--130},
  issn = {00941190},
  urldate = {2026-10-22},
  langid = {english}
}

@article{vickrey1969Am.Econ.Rev.,
  title = {Congestion Theory and Transport Investment},
  author = {Vickrey, William S.},
  year = 1969,
  journal = {The American Economic Review},
  volume = {59},
  number = {2},
  pages = {251--260}
}

@article{rao1988JournaloftheAmericanStatisticalAssociation,
  title = {Resampling {{Inference}} with {{Complex Survey Data}}},
  author = {Rao, J. N. K. and Wu, C. F. J.},
  year = 1988,
  month = mar,
  journal = {Journal of the American Statistical Association},
  volume = {83},
  number = {401},
  pages = {231--241},
  issn = {0162-1459, 1537-274X},
  urldate = {2026-10-22},
  langid = {english}
}

@article{rao1992SurveyMethodology,
  title = {Some Recent Work on Resampling Methods for Complex Surveys},
  author = {Rao, J. N. K. and Wu, C. F. J. and Yue, K.},
  year = 1992,
  journal = {Survey Methodology},
  volume = {18},
  number = {2},
  pages = {209--217}
}

@book{Kotsu2015,
  author = {{Kotsu Shimbunsha}},
  title = {{MY LINE Tokyo Timetable, November 2015 Issue (in Japanese)}},
  year = {2015}
}

@misc{mlit2015railway,
  author = {{Ministry of Land, Infrastructure, Transport and Tourism}},
  title = {National Land Numerical Information: Railway Data},
  year = {2015},
  note = {2015 edition (N02-15), Japan},
  url = {https://nlftp.mlit.go.jp/ksj/gml/datalist/KsjTmplt-N02-v2_3.html},
  shortauthor = {MLIT}
}

@article{henderson1981JournalofUrbanEconomics,
  title = {The Economics of Staggered Work Hours},
  author = {Henderson, J. Vernon},
  year = 1981,
  journal = {Journal of Urban Economics},
  volume = {9},
  number = {3},
  pages = {349--364},
  doi = {10.1016/0094-1190(81)90032-2}
}

@article{takayama2015TransportationResearchPartB:Methodological,
  title = {Bottleneck Congestion and Distribution of Work Start Times: {{The}} Economics of Staggered Work Hours Revisited},
  author = {Takayama, Yuki},
  year = 2015,
  journal = {Transportation Research Part B: Methodological},
  volume = {81},
  pages = {830--847},
  doi = {10.1016/j.trb.2015.07.021}
}

@article{takayama2017JournalofUrbanEconomics,
  title = {Bottleneck Congestion and Residential Location of Heterogeneous Commuters},
  author = {Takayama, Yuki and Kuwahara, Masao},
  year = 2017,
  journal = {Journal of Urban Economics},
  volume = {100},
  pages = {65--79},
  doi = {10.1016/j.jue.2017.05.001}
}

@article{fosgerau2018JournalofUrbanEconomics,
  title = {Vickrey Meets {{Alonso}}: {{Commute}} Scheduling and Congestion in a Monocentric City},
  author = {Fosgerau, Mogens and Kim, Jinwon and Ranjan, Abhishek},
  year = 2018,
  journal = {Journal of Urban Economics},
  volume = {105},
  pages = {40--53},
  doi = {10.1016/j.jue.2018.02.003}
}

@book{efron1993,
  title = {An Introduction to the Bootstrap},
  author = {Efron, Bradley and Tibshirani, Robert},
  year = 1993,
  series = {Monographs on Statistics and Applied Probability},
  number = {57},
  publisher = {Chapman \& Hall},
  address = {New York},
  isbn = {978-0-412-04231-7}
}

\end{document}